\documentclass[prd,a4paper,aps,showpacs,twocolumn,floats]{revtex4}
\usepackage{epsfig}
\usepackage{placeins}
\usepackage{amssymb}
\usepackage{color}
\usepackage{float}

\newcommand{\bk}{{\mathbf k}}

\newcommand{\bV}{{\mathbf V}}

\newcommand{\bx}{{\mathbf x}}

\newcommand{\bn}{{\mathbf n}}
\newcommand{\bv}{{\mathbf v}}

\newcommand{\al}{\alpha}

\newcommand{\de}{\delta}
\newcommand{\De}{\Delta}

\newcommand{\la}{\lambda}
\newcommand{\Om}{\Omega}

\newcommand{\bm}[1]{\mbox{\boldmath $#1$}}

\newcommand{\be}{\begin{equation}}
\newcommand{\ee}{\end{equation}}
\newcommand{\gsim}{\stackrel{>}{\sim}}
\newcommand{\lsim}{\stackrel{<}{\sim}}
\newcommand{\bea}{\begin{eqnarray}}
\newcommand{\eea}{\end{eqnarray}}
\newcommand{\bean}{\begin{eqnarray*}}
\newcommand{\eean}{\end{eqnarray*}}
\newcommand{\dd}{\partial}
\newcommand{\mr}{\mathrm}
\newcommand{\ie}{{\em i.e. }}

\newcommand{\nb}{{\bar n}}
\newcommand{\HH}{{\cal H}}

\begin{document}

\title{What galaxy surveys really measure}

\author{Camille Bonvin}
\email{cbonvin@ast.cam.ac.uk}
\affiliation{Kavli Institute for Cosmology Cambridge and Institute of Astronomy,
Madingley Road, Cambridge CB3 OHA, UK\\
and\\ DAMTP, Centre for Mathematical Sciences, Wilberforce Road, Cambridge CB3 OWA, UK}
\author{Ruth Durrer}
\email{ruth.durrer@physics.unige.ch}
\affiliation{D\'epartement de
Physique Th\'eorique and Center for Astroparticle Physics, Universit\'e de
Gen\`eve, 24 quai Ernest Ansermet, CH--1211 Gen\`eve 4, Switzerland\\
and\\ CEA, SPhT, URA 2306, F-91191 Gif-sur-Yvette, France}

\date{\today}

\begin{abstract}
In this paper we compute the quantity which is truly measured in a large galaxy survey.
We take into account the effects coming from the fact that we actually observe galaxy
redshifts and sky positions and not true spatial positions.
Our calculations are done within linear perturbation theory for both the metric and the 
source velocities but they can be used for non-linear matter power spectra.
We shall see that the complications due to the fact that we only observe on our background 
lightcone and that we do not truly know the distance of the
observed galaxy, but only its redshift is not only an additional difficulty, but even more a new
opportunity for future galaxy surveys.
\end{abstract}


\maketitle

\section{Introduction}
\label{sec:intro}

All the photons which we receive have been emitted on our past light cone.
In cosmology, looking far away always also means looking into the 
past. If the redshift of the objects under consideration is small, $z\ll 1$,
and evolution is relevant only on cosmological time scales, this effect is small 
and may be neglected. However for redshifts of order unity or larger, the fact
that we are not observing a spatial hypersurface but a part of the background
lightcone becomes relevant.

If we observe the large scale distribution of galaxies, we usually
compare the true, observed distribution with the one of an unperturbed 
universe with background density $\bar\rho$ and measure its fluctuations, 
$\de(\bx)=(\rho(\bx)-\bar\rho)/\bar\rho$, where $\bar\rho$ is usually the mean 
observed galaxy density. This is then cast in the power spectrum,
$$ \langle \de(\bk)\de(\bk')\rangle = (2\pi)^3\de(\bk-\bk')P_\de(k) \;,$$
where $\de(\bk)$ is the Fourier transform of $\de(\bx)$ and we assume 
statistical homogeneity and isotropy. 
For small galaxy catalogs one may assume that we measure the density fluctuation today,
$\de(\bx)=\de(\bx,t_0)$, but already for the Sloan Digital Sky Survey (SDSS)
which determines the galaxy distribution out to $z\sim 0.2$ or $0.5$ (for Luminous Red 
Galaxies, LRG's) it is no 
longer a good approximation, to compare the observed power spectrum with the
above defined $P_\de(t_0)$.  Time evolution of $P_\de$ can be taken into account by
multiplying the power spectrum with a growth factor. In addition to this  there is the 
issue of  gauge. The density fluctuation $\de(\bx,t)$ which we calculate in a given
 Friedmann background  is not gauge invariant. It depends
on the background Friedmann universe we compare the observed $\rho(\bx,t)$ with.
This is the cosmological gauge problem~\cite{mybook}.

There are several attempts in the literature to deal with these issues, but they
are so far incomplete.  People have considered  individual
 observational effects like redshift space distortions~\cite{rsd},
 the Alcock-Pacinski effect~\cite{AP} or lensing. A first full treatment is 
attempted in~\cite{Yoo}. In the present paper we shall go beyond this
work and determine the spectrum truly in terms of directly observable quantities.
 We derive gauge invariant expressions which are correct to first order in 
perturbation theory and which are straightforward to compare with observations.
This is an important first step for this problem as the gauge issue is mainly relevant 
 on very large scales, where perturbations are small so that first order 
perturbation theory is justified.

Our results will be most significant for future galaxy catalogs like BOSS~\cite{boss},
DES~\cite{DES}, PanStarrs~\cite{Pan} or, especially Euclid~\cite{Euclid}, but also
an analysis of SLOAN-7~\cite{SDSS} along the lines outlined here is interesting.
\vspace{0.1cm}

{\bf Notation:}  We work with a flat Friedmann background and in conformal time, $t$, such that
$$ ds^2 =a^2(t)\left( -dt^2+\de_{ij}dx^ix^j\right) \,.$$
A photon geodesic in this background which arrives at position $\bx_O$ at time $t_O$ and which
has been emitted at affine 
parameter $\la=0$ at time $t_S$, moving in direction $\bn$ is then given by 
$(x^\mu(\la)) =(\la+t_S, \bx_O+(\la+t_S-t_O)\bn)$. Here $\la = t-t_S = r_S-r$, where $r$ 
is the comoving distance $r=|\bx(\la) -\bx_O|$, hence $dr=-d\la$.  We can of course 
choose $\bx_O=0$.

\section{The matter fluctuation spectrum in redshift space}

In a galaxy redshift survey, the observers measure the number of galaxies in  
direction $\bn$ at redshift $z$, let us call this $N(\bn,z)d\Om_{\bn}dz$. They then 
average over angles to obtain their redshift distribution, 
$\langle N\rangle(z)dz$. From this they can build directly the redshift 
density perturbation \footnote{This is not what is done in practice, where
the observed 'point process' \ie the observed distribution of galaxies is 
compared to a random one with the same redshift distribution (but usually 
many more galaxies to reduce scatter~\cite{SDSS}).}
\ie the perturbation variable
\bea
\de_z(\bn,z) &=& \frac{\rho(\bn,z)-\langle\rho\rangle(z)}{\langle\rho\rangle(z)}
 =\frac{\frac{N(\bn,z)}{V(\bn,z)}-\frac{\langle N\rangle(z)}{V(z)}}
{\frac{\langle N\rangle(z)}{V(z)}}
\nonumber\\
 &=& \frac{N(\bn,z)-\langle N\rangle(z)}{\langle N\rangle(z)}-\frac{\de
 V(\bn,z)}{V(z)}~.
\eea
Here $V(\bn,z)$ is the physical survey volume density per redshift bin, per solid angle.
The volume is also a perturbed quantity since the solid angle of observation 
as well as the redshift bin are distorted between the source and the observer. 
Hence $V(\bn,z)=V(z)+\delta V(\bn,z)$.
The truly observed quantity is the perturbation in the number density of galaxies
\be
\label{Npert}
\frac{N(\bn,z)-\langle N\rangle(z)}{\langle N\rangle(z)}=\de_z(\bn,z)+\frac{\de
 V(\bn,z)}{V(z)} \equiv \De(\bn,z)
\ee
which therefore must be gauge invariant. Actually, as we shall see, both 
$\de_z(\bn,z)$ and $\de V(\bn,z)/V(z)$ are gauge invariant. This is not 
surprising, as we could measure the volume perturbation also with other
tracers than galaxies and it is therefore measurable by itself and hence 
gauge invariant

We neglect biasing in our treatment as we want to keep the expressions as 
model independent as possible. We shall add only some comments on how 
simple linear biasing could be included.

\subsection{Computation of $\de_z(\bn,z)$}

Let us first relate $\de_z(\bn,z)$  to the well known gauge dependent quantity 
$\de(\bx,t)$. For this we note that to first order
\bea
\de_z(\bn,z)&=&\frac{\rho(\bn,z)-\bar \rho(z)}{\bar \rho(z)}=
\frac{\bar{\rho}(\bar{z})+\delta\rho(\bn,z)-\bar\rho(z)}{\bar\rho(z)}\nonumber\\
&=&\frac{\bar{\rho}(z-\de z)+\delta\rho(\bn,z)-\bar\rho(z)}{\bar\rho(z)}
\nonumber \\   \label{e:rhonz}
&=&\frac{\de\rho(\bn,z)}{\bar\rho(\bar z)}-\frac{d\bar \rho}{d \bar z}
\frac{\delta z(\bn,z)}{\bar\rho(\bar z)} .
\eea
Here $\bar z =\bar z(t)$ is the redshift of a background Friedmann universe 
we compare our perturbation with and $\de z$ is the redshift perturbation to 
this universe. Moreover $\rho(\bn,\bar z(t)) =\bar\rho(t) +\de\rho(\bn,t)$,
where the time is obtained by solving the background relation 
$\bar z=\bar z(t)$. Note that $\bar\rho(z) = \bar\rho(\bar z+\de z)$ deviates
to first order from $\bar\rho(\bar z)$. Cleary, both $\de z$ and $\de\rho$ 
depend on the chosen background and are hence gauge dependent. However 
their combination in Eq.~(\ref{e:rhonz}) must turn out to be gauge invariant,
as it is in principle observable.

Let us first compute the redshift in a perturbed Friedman universe with 
metric
\bea
ds^2 &=&a^2(t) \Big[-(1+2A)dt^2 -2B_idtdx^i +  \\
&& +[(1+2H_L)\de_{ij}+ 2H_{Tij} + 2H_{ij}]dx^idx^j\Big]~.\nonumber
\eea
Here $H_{ij}$ is the transverse traceless gravitational wave term and $A$, $B_i$, 
$H_L$ and $H_{Tij}$ are scalar degrees of freedom, two of which can be
removed by gauge transformations. In Fourier space $B_i = -\hat k_i B$ and $H_{Tij} = 
(\hat k_i \hat k_j -\de_{ij}/3)H_T$.  For simplicity, we shall neglect the contribution from gravitational waves 
in the main text. In the appendix we include also these terms. We 
consider a photon emitted from a galaxy, the source, $S$, which is moving in direction 
$\bn$ (hence, to lowest order, it is seen under the direction $-\bn$ from the observer $O$).
We denote the peculiar velocities of the source and observer by
$\bv_S$ and $\bv_O$ . The observer receives the 
photon redshifted by a factor
\be 1+z = \frac{(n\cdot u)_S}{(n\cdot u)_O} \;.
\ee
We solve the equation for the photon geodesic $n = a^{-2}(1+\de n^0,\bn
 +\de\bn)$, where $\bn$ denotes the unperturbed photon direction at the 
observer. Using that $u =a^{-1}(1-A,\bv)$, where $\bv$ is the peculiar velocity, we find
by the same calculation which leads to 
Eq.~(2.228) in~\cite{mybook} (see also~\cite{myrev})
\bea\label{z}
 1+z &=& \frac{a(t_O)}{a(t_S)}\Big\{1 +\Big[H_L+\frac{1}{3}H_T + \bn\cdot\bV +
  \Phi+\Psi\Big]_{t_S}^{t_O}  \nonumber\\
&&  \qquad - \int_{r_S}^{0}(\dot\Phi+\dot\Psi)d\la \Big\} \;.
\eea
The first term is simply $1+\bar z$.
Here $t$ denotes conformal time, $\Psi$ and $\Phi$ are the Bardeen potentials and 
$\bV$ is the gauge invariant velocity perturbation which corresponds to the ordinary 
velocity perturbation in longitudinal gauge. For more details see~\cite{myrev,mybook} 
and Appendix~\ref{app:a}.  In this redshift perturbation the dipole term 
$\bn\cdot\bV(\bx_O,t_O)$ is the only term in the square bracket in (\ref{z}) which 
depends on directions  when evaluated at $\bx_O$. The terms
evaluated at the emission point of course do all depend on $\bn$ via the 
position of the emission point which, to lowest order, is simply 
$\bx_S=\bx_O-\bn(t_O-t(\bar z_S))$.  The integral extends along the unperturbed
photon trajectory from the emission point, where we set $\la=0$, to our position 
where $\la = t_O-t_S = r_S$. Eq. (\ref{z}) implies that the redshift perturbation is
\bea\label{e:dez}
\de z &=& z-\bar{z} =  \nonumber\\
&&  -(1+z)\Big[\big(H_L +\frac{1}{3}H_T + \bn\cdot\bV +
  \Phi+\Psi\big)(\bn,z)   \nonumber \\
&& \qquad +  \int_{0}^{r_S}(\dot\Phi+\dot\Psi)d\la \Big] ,
\eea
where we have neglected the unmeasurable monopole term and 
the dipole term from the observer position. We indicate the 
source position by the direction it is seen under, $-\bn$, and its 
observed redshift $z$.  To lowest order $\bx(\bn,z) = -r_S(z)\bn$.
To obtain the density fluctuation in redshift 
space, we now use that $\frac{d\bar \rho}{d \bar z} = 3\frac{\bar \rho}{1+\bar z}$. 
With this we obtain
\bea
\de_z(\bn,z) &=& D_g(\bn,z) +3(\bV\cdot\bn)(\bn,z) +3(\Psi+\Phi)(\bn,z)
\nonumber \\
&& \qquad + 3\int_{t_S}^{t_O}(\dot\Psi+\dot\Phi)(\bn,z(t))dt \;.  \label{dez2}
\eea 
Here we relate a perturbation variable in direction $\bn$ at redshift $z$
to its unperturbed position and time, $f(\bn,z)=f\left(\bx(\bn,z),t(z)\right)$ and 
overdots are partial derivatives with respect to  $t$, the second argument in $f(\bx,t)$. 

\begin{figure}[!h]
\centerline{\epsfig{figure=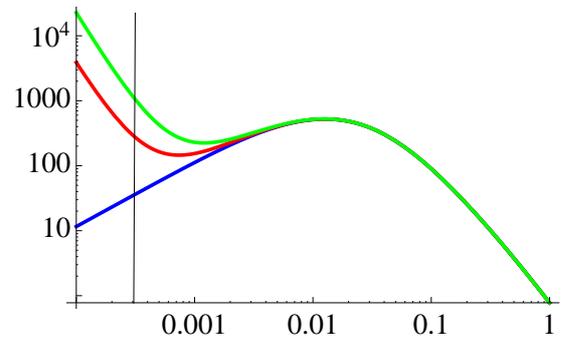,height=4.5cm}}
\caption{ \label{fig:Dens} The (linear) matter  power spectrum on the uniform curvature 
hypersurface (top curve, green), in longitudinal gauge (middle curve, red) and in co-moving 
gauge (bottom curve, blue). }
\end{figure}

$D_g$ is the density fluctuation on the uniform curvature hypersurface. It is related to the 
density fluctuation in co-moving gauge, $D_{cm}$ by~\cite{myrev,mybook}
$$ D_{cm} \equiv D = D_g +3\Phi + 3 k^{-1}\HH V .$$
If we would want to introduce a bias between the matter density and the galaxy density 
it would probably be most physical to assume that both galaxies and dark matter follow the same velocity field as they experience the same gravitational acceleration. We then expect that
biasing should be applied to the density fluctuation in  co-moving gauge, $D_{cm}$, not 
to $D_g$. On small scales such differences are irrelevant but on large scales they do become relevant as becomes clear when considering the (linear) power spectra for the different density fluctuations variables, see Fig.~\ref{fig:Dens}.

\subsection{Volume perturbations}

As next step we compute the volume perturbation $\de V/V$ in Eq.~(\ref{Npert}) which must be
gauge invariant since also $\de_z$ is gauge invariant by itself. This is not surprising as it would 
in principle be a measurable quantity if we would have an 'unbiased tracer' of the volume.

We consider a small volume element at the source position. By this we mean the spatial 
volume seen by a source with 4-velocity $u^\mu$. This is given by 
\be
dV=\sqrt{-g}\;\epsilon_{\mu\nu\al\beta}\;u^\mu \mr{d}x^\nu \mr{d}x^\al \mr{d}x^\beta .
\ee
We want to express the volume element in terms of the polar angles at the observer position, 
$\theta_O$ and $\varphi_O$, and the observed redshift $z$. We have
\bea
dV &=&\sqrt{-g}\;\epsilon_{\mu\nu\al\beta}u^\mu\! \frac{\dd x^\nu}{\dd z} \! 
\frac{\dd x^\al}{\dd \theta_S}\! \frac{\dd x^\beta}{\dd \varphi_S}\! 
\left| \frac{\dd (\theta_S,\varphi_S)}{\dd (\theta_O,\varphi_O)} 
\right|\! \mr{d}z \mr{d}\theta_O\mr{d}\varphi_O \nonumber \\  \label{eq:vol}
&\equiv& v(z,\theta_O,\varphi_O)\mr{d}z\mr{d}\theta_O\mr{d}\varphi_O ~,
\eea
where we have introduced the density $v$ which determines the volume perturbation,
$$ \frac{\de V}{V} =   \frac{v -\bar v}{\bar v} =  \frac{\de v}{\bar v} . $$
$\left| \frac{\dd (\theta_S,\varphi_S)}{\dd (\theta_O,\varphi_O)} \right|$ is the determinant
of the Jacobian of the transformation from the
angles at the source to the angles at the observer. Eq.~(\ref{eq:vol}) is still exact. In a homogeneous and isotropic universe 
geodesics are straight lines and  $\theta_S=\theta_O$ and $\varphi_S=\varphi_O$. In a
perturbed universe the angles at the source are perturbed with respect to the angles at the
observer and we have $\theta_S=\theta_O+\de \theta$ and $\varphi_S=\varphi_O+\de \varphi$. Hence
to first order the Jacobian determinant becomes
\be
\left| \frac{\dd (\theta_S,\varphi_S)}{\dd (\theta_O,\varphi_O)} \right|=1+
\frac{\dd \de \theta}{\dd \theta}+\frac{\dd \de \varphi}{\dd \varphi} .
\ee
Using the expression for the  metric determinant, $\sqrt{-g}=a^4(1+A+3H_L)$ and 
the 4-velocity of the source, $u=\frac{1}{a}(1-A,v^i)$, we find to first order
\bea
v&=&a^3(1+A+3H_L)\Bigg[ \frac{d r}{d z} r^2 \sin\theta_S 
\left(1+\frac{\dd \de \theta}{\dd \theta}+\frac{\dd \de \varphi}{\dd \varphi}\right)\nonumber\\
&-&\left(A\frac{d \bar r}{d\bar z}+v_r\frac{d t}{d z}\right) \bar r^2\sin\theta_O\Bigg] .
\eea
Here $ dr/dz$ is to be understood as the change in comoving distance $r$ with 
redshift along the photon geodesic. At linear order we can write (the distinction 
between $z$ and $\bar z$ is only relevant for background quantities)
\be\label{eq:12}
\frac{d r}{d z}=\frac{d\bar r}{d\bar z}+\frac{d \de r}{d\bar z}-
\frac{d \de z}{d\bar z}\frac{d \bar r}{d \bar z}=\left(\frac{d \bar r}{dt}+
\frac{d \de r}{d\la}-\frac{d\de z}{d\la}\frac{d\bar r}{d \bar z} \right)\frac{dt}{d\bar z} ,
\ee
where we have used that for first order quantities we can set $dt=d\la$ when we
have to take the derivative along the photon geodesic. The last term of Eq.~(\ref{eq:12})
contains the redshift space distortion which will turn out to be the biggest correction to the 
power spectrum.
To lowest order along a photon geodesic $-d\bar r/d\bar z = dt/d\bar z
 =-H^{-1}= -a/\HH$, where $H$ is the physical Hubble 
parameter and $\HH = aH$ is the comoving Hubble parameter. With this the 
volume element becomes 
\bea
\label{Vt}
v&=&\frac{a^4 \bar r^2 \sin \theta_O}{\HH}\Bigg[1+3H_L +
\left( \cot\theta_O+\frac{\dd}{\dd \theta}\right)\de \theta +\frac{\dd \de \varphi}{\dd \varphi}
\nonumber\\
&&- \bv\cdot \bn+\frac{2\de r}{r}-\frac{d\de r}{d\la}+\frac{a}{\HH}\frac{d \de z}{d\la}\Bigg]~.
\eea
 To obtain the fluctuation of $v$ we subtract the unperturbed part $\bar v(z)$. Note, however that we evaluate this at the observed redshift, $z=\bar z +\de z$. Hence
$$ \bar v(z) = \bar v(\bar z) + \frac{d\bar v}{d\bar z}\de z . $$
From the unperturbed expression with $a=1/(\bar z+1)$,
\be
\bar v(\bar z) = \frac{\sin\theta_O \bar r^2}{(1+\bar z)^4 \HH}
\ee
one infers
\be
\label{dVz}
\frac{d\bar{v}}{d \bar z}=\bar v(\bar z) 
\left(-4+\frac{2}{\bar r_S\HH}+\frac{\dot{\HH}}{\HH^2} \right)\frac{1}{1+\bar z}~.
\ee
Combining Eq.~(\ref{Vt}) and (\ref{dVz}) we obtain for the perturbation of the 
volume element 
\bea
\lefteqn{\frac{\de v}{\bar v}(\bn,z) = \frac{v(z) -\bar v(z)}{\bar v(z)}  = }  \\
&& \hspace*{-2mm}3H_L+\left( \cot\theta_O+\frac{\dd}{\dd \theta}\right)\de \theta +
\frac{\dd \de \varphi}{\dd \varphi}- \bv\cdot \bn+\frac{2\de r}{r}\nonumber\\
&&\hspace*{-2mm}-\frac{d\de r}{d\la}+\frac{1}{\HH(1+\bar z)}\frac{d \de z}{d\la}
-\left(-4+\frac{2}{\bar r\HH}+\frac{\dot{\HH}}{\HH^2} \right)\frac{\de z}{1+\bar z} .\nonumber
\label{e:volpert}
\eea

In order to express these quantities in terms of the perturbed metric and the 
peculiar velocity of observer and emitter, we need to compute the deviation vector that 
relates the perturbed geodesic to the unperturbed one 
$\de x^\mu(\la)=x^\mu(\la)-\bar x^\mu(\la)$.
Here we give only the main steps. More details on the derivation can be 
found in the appendix. We use
\be
\frac{d x^\mu}{dt}=\frac{d x^\mu}{d\la}\frac{d\la}{dt}=\frac{n^\mu}{n^0}
\ee 
which leads to
\bea
x^0(t_S)&=&-(t_O-t_S)  = r_S \hspace{0.3cm} \mbox{at every order}\\
x^i(t_S)&=&-(t_O-t_S)\bar n^i-\int_{0}^{r_S} d\la(\de n^i - \bar n^i\de n^0)  
\eea
to first order.

In the following we neglect perturbations at the observer position since, 
as already mentioned, they give rise only to unmeasurable monopole term or 
a dipole term. Using the null geodesic equation for $n^\mu$ we find
\bea
\de x^i(t_S)&=&+\int_{0}^{r_S} d\la \Big( h_{\al i}\nb^\al + h_{0\al}\nb^i\nb^\al \Big)\\
&+&\frac{1}{2}\int_{0}^{r_S} d\la(r_S-r) \Big( h_{\al\beta,i}+
\dot{h}_{\al\beta}\nb^i\Big)\nb^\al\nb^\beta\nonumber ,
\eea
where $r(\la) = \la$. From this we obtain
\bea
\de r &\equiv& \de x^i e_{r i}=-\de x^i \nb_i=-\frac{1}{2}\int_{0}^{r_S} d\la h_{\al\beta}\nb^\al\nb^\beta
\nonumber \\    \label{dr_gi}
&=&\int_{0}^{r_S} \hspace{-3mm}d\la (\Phi+\Psi)+\frac{B}{k}+
\frac{1}{k^2}\!\left(\!\frac{dH_T}{d\la}-2\dot{H}_T\! \right) . \eea
We also use that 
$\nb^i\dd_i+\dd_t=\frac{d}{d\la}=\frac{d}{dt}$  and $r_S=t_O-t_S$ to lowest order.
For the derivative of $\de r$ we obtain
\be
\label{drdt_gi}
\frac{d\de r}{d\la}=-(\Phi+\Psi)+\frac{1}{k}\frac{dB}{d\la}+
\frac{1}{k^2}\left(\frac{d^2H_T}{d\la^2}-2\frac{d\dot{H}_T}{d\la} \right) .
\ee
Similarly we find for the perturbed angles 
\bea
\de \theta&\equiv&\frac{\de x^ie_{\theta i}}{r_S}=\frac{1}{r_S}\int_{0}^{r_S} d\la \nonumber\\
&&\times\Big(h_{\al i}\nb^\al e_\theta^i
+\frac{1}{2}(r_S-r)h_{\al\beta,i}e_\theta^i\nb^\al\nb^\beta \Big) \label{dtheta}~,\\
\de \varphi&\equiv&\frac{\de x^ie_{\varphi i}}{r_S\sin\theta_O}=
\frac{1}{r_S\sin\theta_O}\int_{0}^{r_S} d\la\nonumber\\
&& \times\Big(h_{\al i}\nb^\al e_\varphi^i
+\frac{1}{2}(r_S-r)h_{\al\beta,i}e_\varphi^i\nb^\al\nb^\beta \Big).
\eea
We have used that $\nb^i e_{\theta i}=\nb^i e_{\varphi i}=0$.
The second term of the integral in Eq.~(\ref{dtheta}) can be rewritten as 
\bea
h_{\al\beta,i}e_\theta^i\nb^\al\nb^\beta&=&\frac{1}{r}\dd_\theta(h_{\al\beta})\nb^\al\nb^\beta\\
&=&\frac{1}{r}\Big[\dd_\theta(h_{\al\beta}\nb^\al\nb^\beta)-h_{\al\beta}\dd_\theta(\nb^\al\nb^\beta)\Big]~,\nonumber
\eea
where $\dd_\theta\nb^\al=-e_\theta^i\de_{i\al}$, and analogously for $\varphi$. The angular 
contribution to the volume then reads
\bea
\lefteqn{(\cot\theta+\dd_\theta) \de\theta +\dd_\varphi\de\varphi=
\int_{0}^{r_S} d\la\frac{(r_S-r)}{2r_Sr}}\nonumber\\
&& \hspace{-0.1cm}\times \Big[ \cot\theta\dd_\theta+\dd^2_\theta+\frac{1}{\sin^2\theta} 
\dd^2_\varphi \Big]h_{\al\beta}\nb^\al\nb^\beta \nonumber\\
&& \hspace{-1cm} +\int_{0}^{r_S} d\la \frac{1}{r}\Big[ (\cot\theta+\dd_\theta)h_{i\al}e_\theta^i\nb^\al +\frac{\dd_\varphi}{\sin\theta}h_{i\al}e_\varphi^i\nb^\al\Big]\nonumber \\
\label{angle}
&=& \frac{-1}{r_S}\int_{0}^{r_S} d\la\frac{(r_S-r)}{r}\Delta_\Om(\Phi+\Psi) \nonumber \\
  && \qquad - \frac{\Delta_\Om H_T(t_S)}{k^2r_S^2}~,
\label{angle_gi}
\eea
where $\Delta_\Omega$ denotes the angular part of the Laplacian 
\be
\Delta_\Om\equiv\Big( \cot\theta\dd_\theta+\dd^2_\theta+\frac{1}{\sin^2\theta} \dd^2_\varphi\Big)~.
\ee
It is interesting to note that the angular part of the volume perturbation is not a gauge-invariant 
quantity by itself. If $H_T\neq 0$ the angular and radial directions are mixed in a non-trivial way.
This is not really surprising since the angular volume distortion is not a measurable quantity by itself.
On the other hand, the convergence $\kappa$ (or the magnification $\mu$) that are observable, contain
in addition to the angular volume distortion other perturbations (see~\cite{bonvin},
\cite{bernardeau}) and are consequently gauge invariant.

The redshift contribution to the volume perturbation is obtained by differentiating Eq.~(\ref{e:dez}).
\bea
\label{dzdt_gi}
\lefteqn{\frac{1}{\HH(1+\bar z)}\frac{d\de z}{d\la}=}\\
&&\Phi+\Psi +H_L+\frac{H_T}{3}+\bV\cdot\bn
+\int_{0}^{r_S} d\la(\dot\Phi+\dot\Psi)\nonumber\\
&&-\frac{1}{\HH}\left(\bar n^i\dd_i(\Phi+\Psi)+\frac{dH_L}{d\la}+\frac{1}{3}\frac{dH_T}{d\la}
+\frac{d(\bV\cdot\bn)}{d\la} \right) . \nonumber
\eea
Putting everything together we find after several integrations by part and a total Laplacian 
of $H_T$ which cancels a factor $1/k^2$, the following expression for the volume density perturbation
\bea
\lefteqn{\frac{\de v}{v}=-2(\Psi+\Phi) -4\bV\cdot\bn +\frac{1}{\HH}
\left[\dot\Phi+\dd_r\Psi-\frac{d(\bV\cdot\bn)}{d\la}\right]} \nonumber\\
&&+\left(\frac{\dot{\HH}}{\HH^2}+\frac{2}{r_S\HH}\right)
\left(\Psi+\bV\cdot\bn+ \int_{0}^{r_S} d\la(\dot{\Phi}+\dot{\Psi})\right)\nonumber\\
&&-3\int_{0}^{r_S} d\la(\dot{\Phi}+\dot{\Psi})+ \frac{2}{r_S}\int_{0}^{r_S} d\la (\Phi+\Psi)\nonumber\\
&&- \frac{1}{r_S}\int_{0}^{r_S} d\la\frac{r_S-r}{r}
\Delta_\Om(\Phi+\Psi)~.  \label{dev}
\eea
Here and in the following, the functions without argument are to be evaluated at the 
source position $\bx_S = \bx_O -\bn(t_O-t_S)$ and at the source time $t_S$.
More details of the derivation of this result are given in the appendix.

Adding the results (\ref{dez2}) and (\ref{dev}) we obtain  the galaxy number density 
fluctuation in redshift space as defined in Eq.~(\ref{Npert})
\bea
\De(\bn,z) &=& D_g + \Phi + \Psi + \frac{1}{\HH}
\left[\dot\Phi+\dd_r(\bV\cdot\bn)\right] \nonumber \\  && \hspace{-1.9cm}+ 
 \left(\frac{\dot{\HH}}{\HH^2}+\frac{2}{r_S\HH}\right)\left(\Psi+\bV\cdot\bn+ 
 \int_0^{r_S}\hspace{-3mm}d\la(\dot{\Phi}+\dot{\Psi})\right) 
  \nonumber \\  &&   \label{Dez}
 \hspace{-0.8cm} +\frac{1}{r_S}\int_0^{r_S}\hspace{-3mm}d\la \left[2 - \frac{r_S-r}{r}\Delta_\Om\right] (\Phi+\Psi) .
\eea
Here we have used that also pressureless matter moves along geodesics so that
$$ \bn\cdot\dot\bV +\HH\bn\cdot\bV -\dd_r\Psi =0\,.$$
Equation~(\ref{Dez}) together with (\ref{dez2}) and (\ref{dev}) are our first main result.

The first term in~(\ref{Dez}) is the gauge invariant density fluctuation. $D_g$ is the density fluctuation
in the flat slicing. It is related to the density perturbation in Newtonian gauge by 
$D_g = D_s -3\Phi$. In terms of $D_s$ the first three contributions combine to 
$ D_g + \Phi + \Psi =  D_s - 2\Phi + \Psi$.

The term $\HH^{-1}\dd_r(\bn\cdot\bV)$ is the redshift space distortion. As we shall see
in the next section, this is the largest single correction on intermediate scales.
The second line comes from the redshift perturbation of the volume. It contains a 
Doppler term and the ordinary and integrated Sachs-Wolfe terms.
The third line represents the radial and angular volume distortions.  The
second term in the integral on the third line is especially relevant on large scales, 
it is the lensing distortion.

\section{The angular power spectrum of the galaxy density fluctuations}
For fixed redshift $\De(z,\bn)$ is a function on the sphere and it is most natural to expand it
in spherical harmonics. Let us do this with the result (\ref{Dez})
\be
\De(\bn,z) =\sum_{\ell m}a_{\ell m}(z)Y_{\ell m}(\bn) , \quad C_\ell (z) = \langle |a_{\ell m}|^2 \rangle .
\ee
The coefficients $a_{\ell m}(z)$ are given by
\be
a_{\ell m}(z) = \int d\Om_{\bn}Y_{\ell m}^*(\bn)\De(\bn,z).
\ee
The star indicates complex conjugation.

The different terms in $\De(\bn,z)$ are either a perturbation variable 
evaluated at the source position or an integral of a perturbation variable 
over the unperturbed photon trajectory. Let us first consider a 
contribution from a perturbation variable at the source position, e.g. 
$\Psi$. We want to relate the $C_\ell(z)$ 
 spectra to the usual power spectrum $P_\Psi(k,t)$ which is defined by
 $$
 \langle \Psi(\bk,t)\Psi^*(\bk',t)\rangle  =(2\pi)^3\de(\bk-\bk')P_\Psi(k,t) .
$$
The delta function and the fact that $P_\Psi$ depends only on the modulus 
of $\bk$, $k\equiv |\bk|$, are a consequence of statistical homogeneity and isotropy.
Expressing  $\Psi$ in terms of its Fourier transform,
$$ \Psi(\bx,t) = \frac{1}{(2\pi)^3}\int d^3k \Psi(\bk,t)e^{-i(\bk\cdot\bx)} , $$
a short calculation (see e.g.~\cite{mybook}) gives
\be
a^\Psi_{\ell m}(z_S) = \frac{i^{\ell}}{2\pi^2}\int d^3k j_{\ell}(kr_S)\Psi(\bk,t_S)Y^*_{\ell m}(\hat{\bk}) .
\ee
Here $j_\ell$ is the spherical Bessel function of order $\ell$, see~\cite{AS}.
Correspondingly the contribution from an integral $\int_{0}^{r_S}f(\bx(\la),t(\la)) d\la$ becomes
\be
a^{\int\! \! f}_{\ell m}(z_S) = \frac{i^{\ell}}{2\pi^2}\int_{0}^{r_S}\hspace{-2mm}
d\la\int d^3k j_{\ell}(k\la)f(\bk,t)Y^*_{\ell m}(\hat{\bk}) .
\ee
For a velocity term $\bV\cdot\bn$ we use that $\bV(\bk,t) = i\hat\bk V$, so that
$ \bV\cdot\bn\exp[i(\bk\cdot\bn)r] = V\dd_{kr}\exp[i(\bk\cdot\bn)r]$. 
With this one obtains
\be
a^{\bV\bn }_{\ell m}(z_S) = \frac{i^{\ell}}{2\pi^2}\int d^3k j'_{\ell}(kr_S)V(\bk,t)Y^*_{\ell m}(\hat{\bk}) .
\ee
The prime in $j_\ell$ denotes derivation w.r.t the argument.  Finally, for the  
redshift space distortion, $\dd_r(\bV\cdot\bn) = -\bn\cdot{\bm{\nabla}}(\bV\cdot\bn)$
we have to use the above identity twice and arrive at
\be
a^{\dd_r(\bV\bn) }_{\ell m}(z_S) = \frac{i^{\ell}}{2\pi^2}\int d^3k j''_{\ell}(kr_S)k^{-1}
V(\bk,t)Y^*_{\ell m}(\hat{\bk}) .
\ee

One can now write down the $C_\ell(z)$'s for one's theory of choice for the background 
and the perturbations, {\em e.g.} for modified gravity or a quintessence model. 

So far the derivation has been completely general.  We have not used Einstein's equation. The 
only assumptions are that galaxies follow the distribution of matter which
is made out of non-relativistic particles which move along geodesics, and that 
photons move along null geodesics.

To proceed further, we have to be more specific. Here we just study the simplest model
of purely {\em scalar adiabatic perturbations},  which have been generated
at some early time in the past (e.g. inflation). If there are more e.g.
isocurvature modes present, the subsequent calculation has to be repeated for them.

In the case of one adiabatic mode, all the perturbation variables are given by transfer functions
from some initial random variable that we take to be the Bardeen potential $\Psi$.
Hence
\bea
\Psi(\bk,t) &=& T_\Psi(k,t)\Psi_{\rm in}(\bk) \\
\Phi(\bk,t) &=& T_\Phi(k,t)\Psi_{\rm in}(\bk) \\
D_g(\bk,t) &=& T_D(k,t)\Psi_{\rm in}(\bk) \\
V(\bk,t) &=& T_V(k,t)\Psi_{\rm in}(\bk) .
\eea   
The transfer functions $T_{\small\bullet}$ depend on the matter content 
and the evolution history of the Universe and on the theory of gravity 
which relates matter and metric degrees of freedom. What is important for
us is that they are deterministic functions and do not depend on directions
of $\bk$. We characterize the initial power spectrum by a spectral 
index, $n$, and an amplitude, $A$, 
\be 
k^3\langle \Psi_{\rm in}(\bk)\Psi^*_{\rm in}(\bk')\rangle  =(2\pi)^3\de(\bk-\bk')A(kt_O)^{n-1} .
\ee
We have introduced present time, $t_O$, in order to keep $A$ 
dimensionless. From the CMB observations we know that it is of the order 
of $A\sim 10^{-8}$. With these identifications we can now relate
$C_\ell(z)$ to the initial power spectrum $A (kt_O)^{n-1}$. Inserting the 
above in expression (\ref{Dez}) a short calculation gives
\be
C_\ell(z_S) = \frac{2A}{\pi}\int\frac{dk}{k}(kt_O)^{n-1} \left|F_\ell(k,z_S)\right|^2
\ee
with
\onecolumngrid
\bea
F_\ell(k,z_S) &=& j_\ell(kr_S)\!\left[\!T_D +\left(\!1\! +\!\frac{\dot \HH}{\HH^2} \!+
 \!\frac{2}{r_S\HH}\!\right)T_\Psi   \!+ \! T_\Phi \!+ \! \frac{1}{\HH}\dot T_\Phi \!\right] 
+ j_\ell'(kr_S)\left(\!\frac{\dot \HH}{\HH^2} \!
+\! \frac{2}{r_S\HH} \right)T_V  \!+\! \frac{k}{\HH}T_V j_\ell''(kr_S) 
   \nonumber \\  && 
  \hspace{-1.2cm} + \frac{1}{r_S}\int_{0}^{r_S}j_\ell(k\la)\left(2 +\frac{r_S-\la}{\la}\ell(\ell+1)\right)
   (T_\Psi + T_\Phi)d\la   + \left(\frac{\dot \HH}{\HH^2} + \frac{2}{r_S\HH}\right)   
\int_{0}^{r_S}j_\ell(k\la)(\dot T_\Psi +\dot T_\Phi)d\la
  .    \label{Flkz}
\eea
\twocolumngrid
Here $r_S=t_O-t_S$ is the source position .

We now  evaluate and compare the amplitude of different terms in a $\Lambda$CDM universe. 
Rather than entering in a precise numerical evaluation, we estimate the terms by using  approximations for the transfer functions. This will help us to gain insight in the importance of 
the different terms. We plan to do a full numerical evaluation which can be used to estimate cosmological parameters in future work. 

From the first order Einstein equations, neglecting anisotropic stresses from neutrinos, we can 
relate the transfer functions $T_D, T_V$ and $T_\Phi$ to
$T_\Psi$. We find
\bea
T_\Phi&=&T_\Psi    \label{TPhi}\\
T_D&=& -\frac{2 a}{3\Om_m}\left(\frac{k}{\HH_0}\right)^2T_\Psi-3T_\Psi-3\frac{\HH}{k}T_V\label{TD}\\
T_V&=&\frac{2 a}{3\Om_m}\frac{k}{\HH_0^2}\left(\HH T_\Psi+\dot{T}_\Psi \right) \label{TV}
\eea
Using the notation of \cite{dod} (see also \cite{chris}), we decompose the transfer 
function $T_\Psi(k,t)$ into a growth rate $D_1(a)$ and a time independent transfer 
function $T(k)$ such that 
\be\label{TPsi}
T_\Psi(k,t)=\frac{9}{10}\frac{D_1(a)}{a}T(k), 
\ee
and we use 
CAMBcode~\cite{CAMB}  to compute $T(k)$.
The amplitude of the power spectrum can be expressed as~\cite{dod} 
$A=\frac{50\pi^2}{9}\delta_H^2\left(\frac{\Om_m}{D_1(a=1)} \right)^2$.
We choose $\Om_m=0.24$, $\Om_\Lambda=0.76$ and $\sigma_8=0.75$ leading to $\delta_H=5.6\cdot 10^{-5}$.

\subsection{The transversal power spectrum}

\begin{figure}[ht]
\centerline{\epsfig{figure=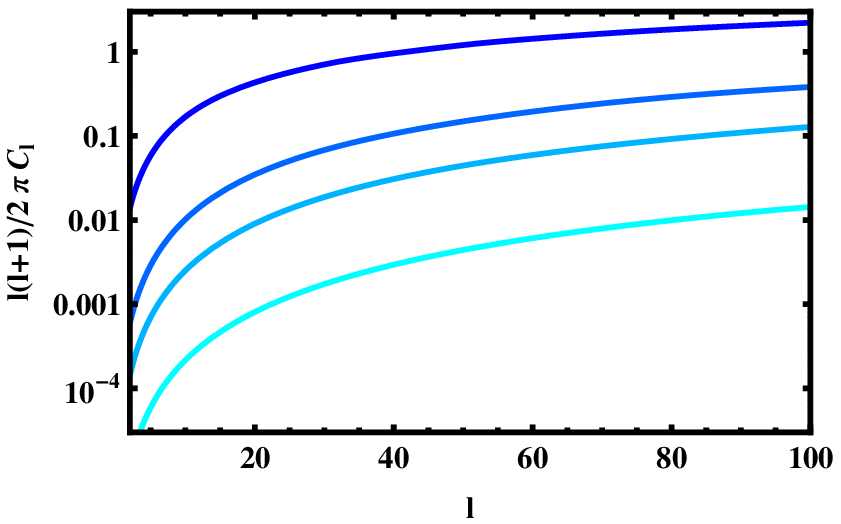,height=5.3cm}}
\centerline{\epsfig{figure=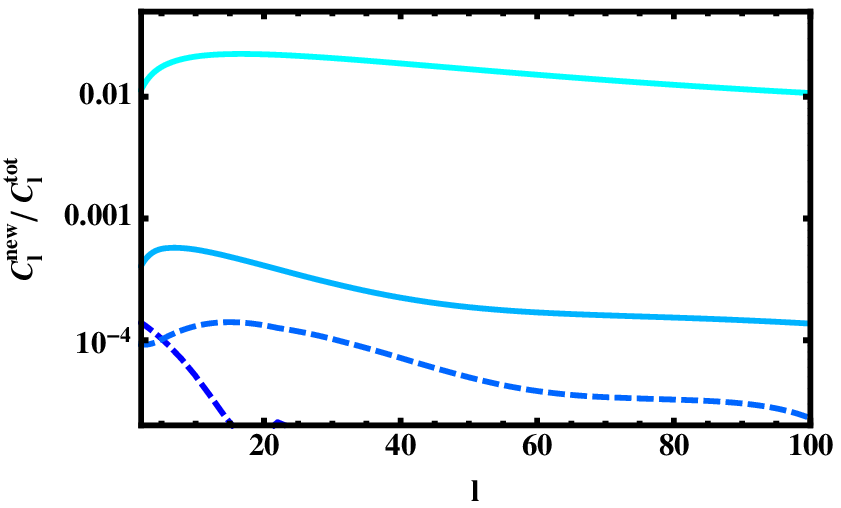,height=5.3cm}}
\caption{ \label{f:tot} Top panel: The transversal power spectrum at (from top to bottom) $z_S=0.1$, $z_S=0.5$, 
$z_S=1$ and $z_S=3$. \\
Bottom panel: The ratio between the new contributions (lensing+potential) and the total angular power spectrum 
at (from top to bottom) $z_S=3$, $z_S=1$, $z_S=0.5$ and $z_S=0.1$. 
Solid lines denote positive contributions whereas dashed lines denote negative contributions.}
\end{figure}

Let us first determine the $C_\ell$'s at fixed redshift. They provide the transversal power spectrum,
i.e. correlations on the sphere normal to the observer directions. Of course for the intrinsic density
fluctuations these are not different from correlations in any other direction, but observational effects on them are different. E.g., since we can only observe on the background lightcone, we can only
see fluctuations on this sphere at the same time but not fluctuations which have a 
different radial distance from us. On the other hand, in general the same redshift does 
of course not imply the same look-back time, since both these quantities are perturbed in different ways.

\begin{figure}[!h]
\centerline{\epsfig{figure=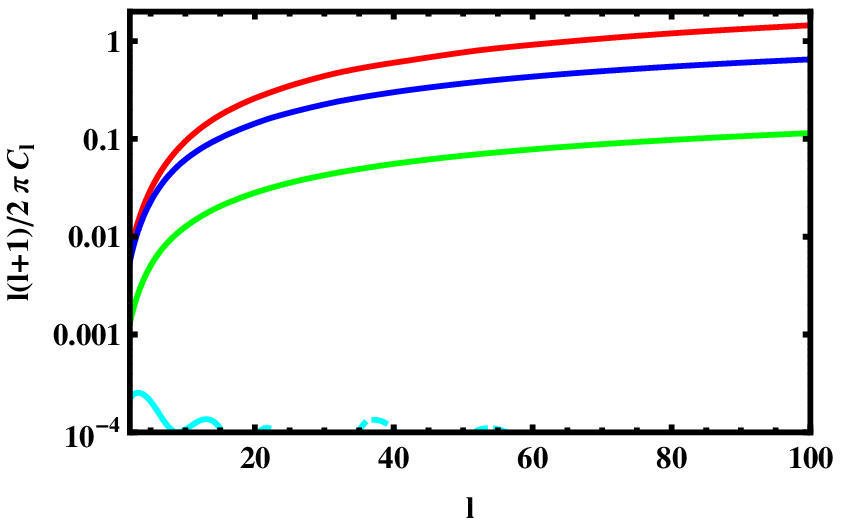,height=5cm}}
\centerline{\epsfig{figure=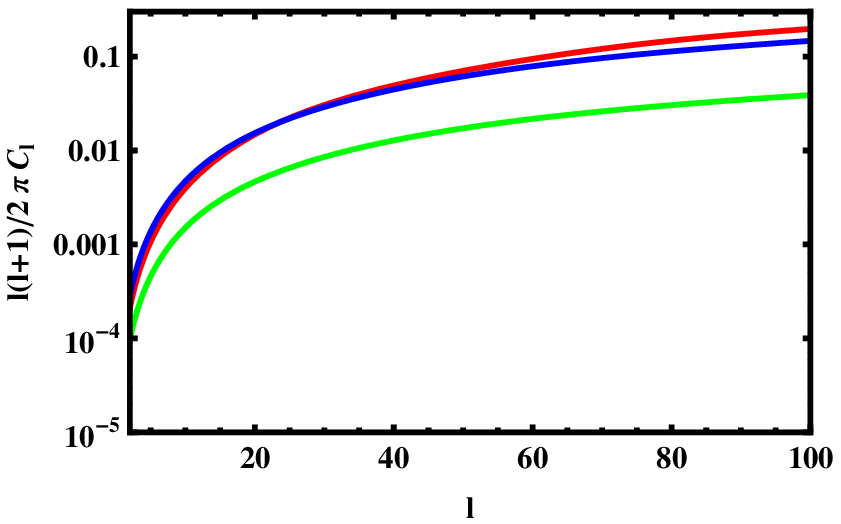,height=5cm}}
\centerline{\epsfig{figure=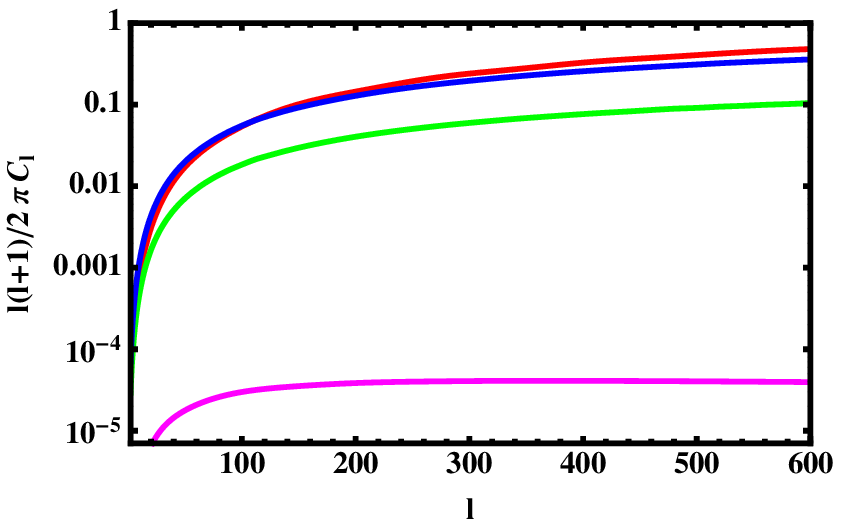,height=5cm}}
\centerline{\epsfig{figure=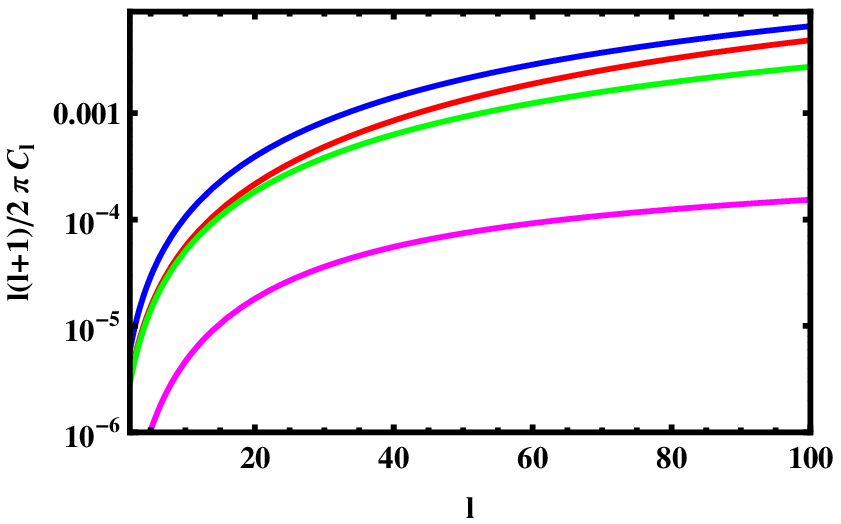,height=5cm}}
\caption{ \label{f:z01to3} The  dominant terms at redshifts (from top to bottom) $z_S~=~0.1,~0.5,~1$ and $3$: density (red),  redshift space distortion (green), 
the correlation of density with redshift space distortion (blue), lensing (magenta), 
Doppler (cyan), see Table~\ref{t:color}. The potential terms
are too small to appear on our log-plot.}
\end{figure}

\begin{figure}[ht]
\centerline{\epsfig{figure=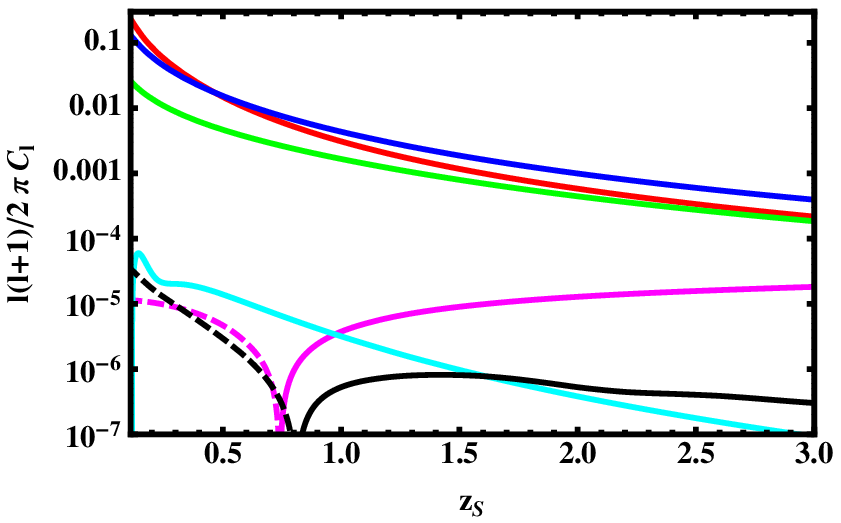,height=5cm}}
\centerline{\epsfig{figure=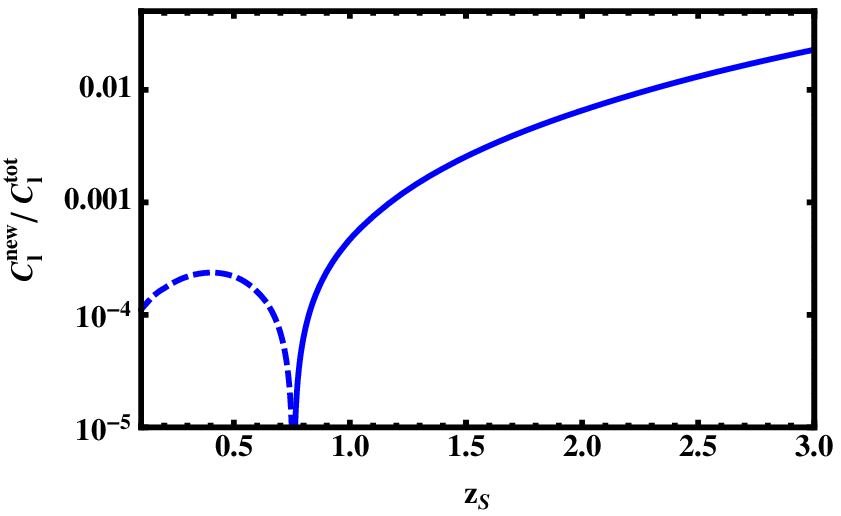,height=5cm}}
\caption{ \label{f:zl20} Top panel: The various terms as a function of $z_S$ for fixed value of
$\ell=20$: density (red), redshift space distortion (green), the correlation of density
with redshift space distortion (blue), lensing (magenta), Doppler (cyan) and potential 
(black), see Table~\ref{t:color}. Solid lines denote positive contributions whereas dashed lines denote negative contributions.\\
Bottom panel: The ratio between the new contributions (lensing+potential) and the total angular power spectrum
as a function of $z_S$ for fixed value of $\ell=20$.}
\end{figure}

\begin{table}
\begin{tabular}{|l|c|c|}
\hline
Density & $D$ & {\color{red}\bf red} \\ \hline
redshift space  & & \\
distortion & $\HH^{-1}\dd_r(\bV\cdot\bn)$& {\color{green}\bf green} \\ \hline
lensing & $\frac{-1}{r_S}\int_0^{r_S}\hspace{-1mm}d\la\frac{r_S-r}{r}\Delta_\Om(\Phi+\Psi)$&
  {\color{magenta}\bf magenta}\\ \hline
  correlation & $2D\HH^{-1}\dd_r(\bV\cdot\bn)$& {\color{blue}\bf blue} \\ \hline
Doppler &  $\left(\frac{\dot{\HH}}{\HH^2}+\frac{2}{r_S\HH}\right)\bV\cdot\bn$& 
 {\color{cyan}\bf cyan}\\ \hline
potential &$ \Psi - 2\Phi +\frac{2}{r_S}\int_0^{r_S}\hspace{-1mm}d\la(\Phi+\Psi)+ $ & \\  & $\left(\frac{\dot{\HH}}{\HH^2}+\frac{2}{r_S\HH}\right)\left[\Psi+  \int_0^{r_S}\hspace{-1mm}d\la(\dot{\Phi}+\dot{\Psi})\right]  $
& \\  & $
 -\frac{2a}{\Omega_m}\left( \frac{\HH}{\HH_0}\right)^2\left(\Psi+\frac{\dot{\Phi}}{\HH} \right)$ & {\bf  black}\\
\hline
\end{tabular}
\caption{\label{t:color}The color coding the different terms of Eq.~(\ref{Dez}) in the 
angular power spectrum of $\De(\bn,z)$ as shown in 
Figs.~\ref{f:z01to3} to~\ref{f:z01to3l20} and \ref{f:win}, \ref{f:zwinl20}, \ref{f:winz01to3l20}. 
In addition to the term given in the second column, all its correlations with the terms 
in the lines above are also included. Only the most dominant correlation
between density and redshift space distortion is shown separately in blue. In 
Figs.~\ref{f:z01to3l20}  and  \ref{f:winz01to3l20} the 'standard terms', i.e. the top three 
lines, are represented together as the blue line. }
\end{table}

In Fig.~\ref{f:tot} (top panel) we show the total transversal power spectrum at redshifts $z_S=0.1,~0.5,~1$
and $3$. Note that the amplitude of the linear power spectrum from $z_S=0.1$ to $z_S=0.5$ is reduced by a factor 6 at $\ell \sim 100$ and by a factor 20 at $\ell\lsim 10$. This comes 
from the following fact: the transversal power spectrum is dominated by the density fluctuation
and the redshift space distortion which are proportional to integrals of the form
 $$ 
\int \frac{dk}{k}\left(\frac{k}{\HH_0} \right)^4T^2(k)j_\ell^2(kr_S) \,.
$$
At $x=kr_S=\ell$, this term goes like $\ell^4$ and it is therefore expected to dominate 
at large $\ell$. However, since for a constant transfer function, this integral would diverge,
it is dominated by the maximum of the transfer function which is roughly at
$k_{\rm eq}$. Since for $z\gsim 0.5$, $k_{\rm eq}r_S \gsim \ell$, 
$j_\ell^2(k_{\rm eq}r_S) \propto 1/(k_{\rm eq}r_S)^2$ which therefore decreases 
like $1/r_S^2$.  Already this 
simple observation tells us that the amplitude of the transversal power spectrum at
different redshifts might 
offer a possibility to constrain $r_S(z)$ and the growth factor, which both depend on 
cosmological parameters in different ways. On the other hand, this is complicated by 
non-linear effects and biasing which are not accounted for in this work.

The different contributions to the power spectrum at different 
redshifts are shown in more detail in Fig.~\ref{f:z01to3}.  For 
$z_S=1$, we show the spectrum up to $\ell=600$ while for the other 
redshifts we stop at $\ell = 100$ beyond which the structure does not change anymore.
We denote by $D$ the density term in co-moving gauge, 
$$D = D_g + 3\Phi + 3\frac{\HH}{k}V \,,$$
by $z$ the redshift space distortion, 
by $L$ the lensing term, by $V$ the Doppler terms and by $\Psi$ the gravitational potential 
terms (see Table~\ref{t:color} for a definition of each term). $C^{DD}_\ell$ represents for example the contribution from the density term alone 
and $C^{Dz}_\ell$ the correlation between the density and redshift space distortion.
Except from the correlation between the density and redshift space distortion that we 
represent individually, we include the correlations with the smaller contribution. Note 
that usually the correlations between lensing, Doppler and gravitational potential are 
negligible and, except when explicitly specified we do neglect them.
Therefore, when we plot for example the lensing term (magenta), it
contains $C^{LL}_\ell+2C^{LD}_\ell+2C^{Lz}_\ell$.  
The formulae for the dominant $C_\ell$'s are given in Appendix~\ref{app:Cls}.

\begin{figure}[!h]
\centerline{\epsfig{figure=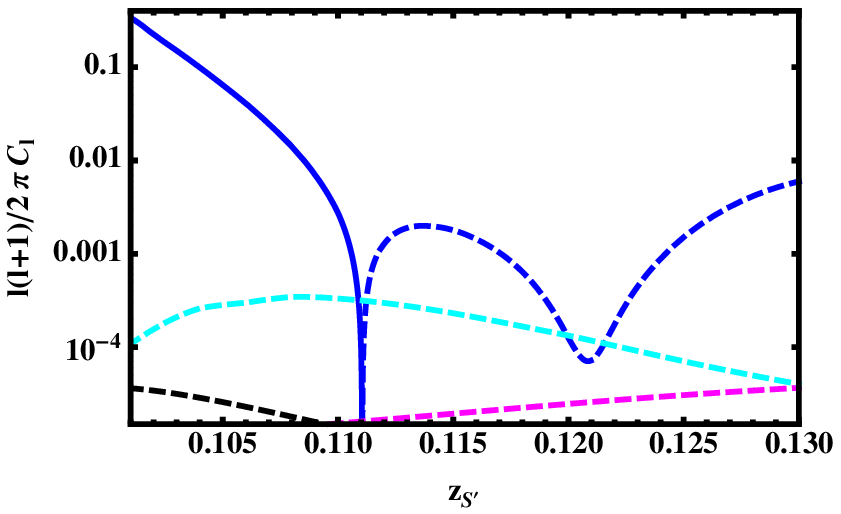,height=5cm}}
\centerline{\epsfig{figure=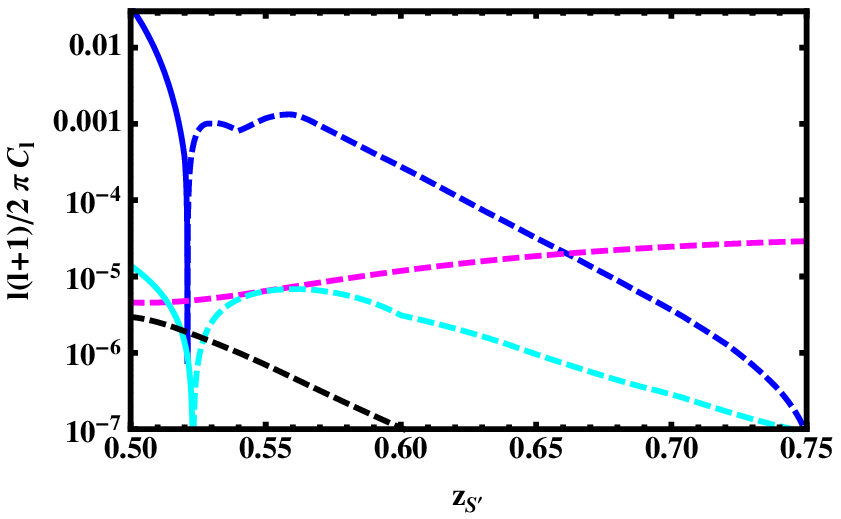,height=5cm}}
\centerline{\epsfig{figure=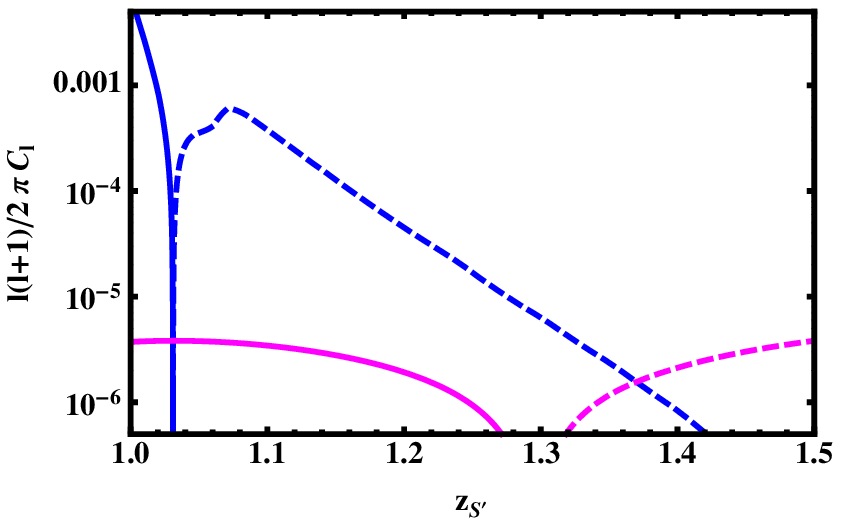,height=5cm}}
\centerline{\epsfig{figure=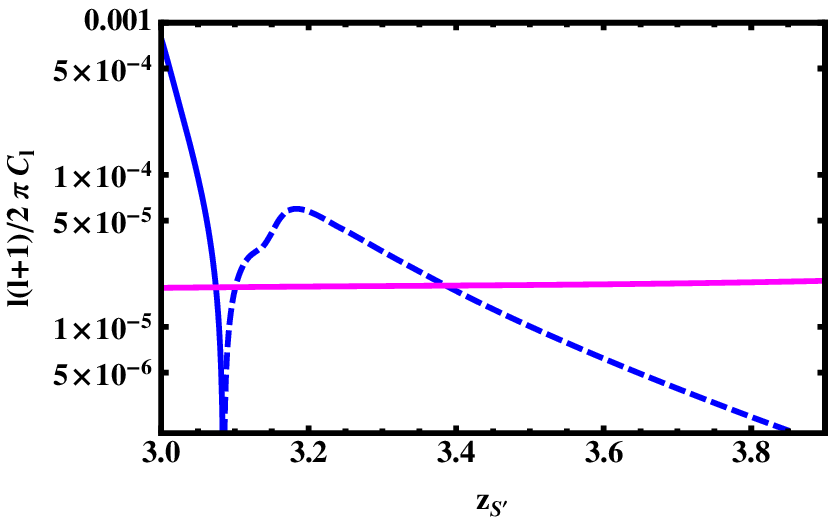,height=5cm}}
\caption{ \label{f:z01to3l20} Different terms of $C_\ell(z_S,z_{S'})$ at $\ell=20$ for redshifts (from top to bottom)
$z_S=0.1,~0.5,~1$ and $3$,
plotted as a function of $z_{S'}$: standard term, \ie $C^{DD}_\ell+C^{zz}_\ell+2C^{Dz}_\ell$ (blue), 
lensing (magenta), Doppler (cyan), potential (black), see Table~\ref{t:color}.
Solid lines denote positive contributions whereas dashed lines denote negative contributions.}
\end{figure}

\begin{figure}[ht]
\centerline{\epsfig{figure=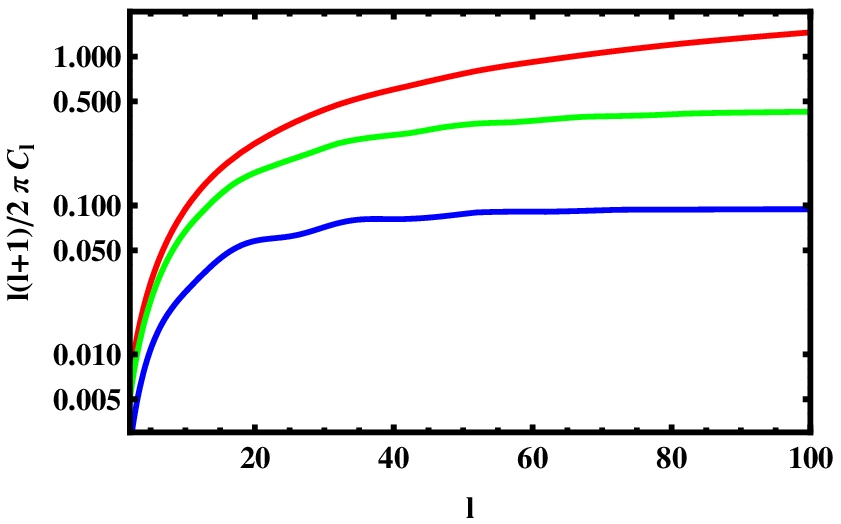,height=5cm}}
\centerline{\epsfig{figure=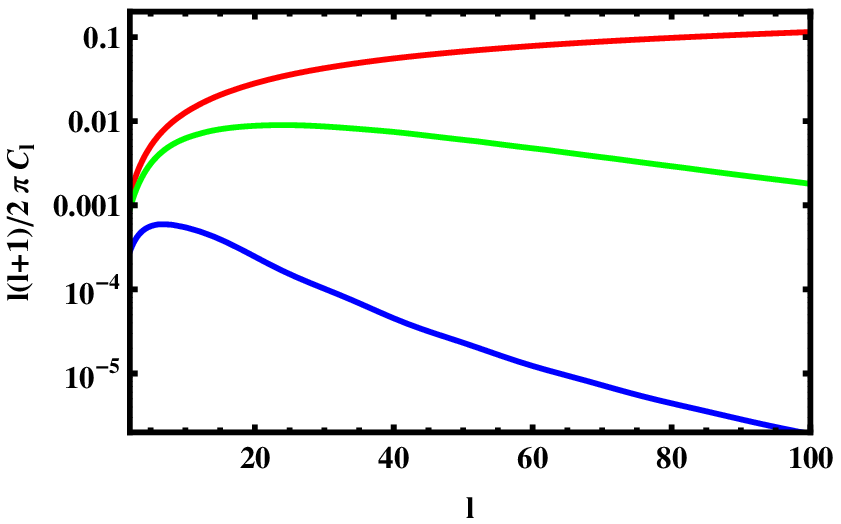,height=5cm}}
\centerline{\epsfig{figure=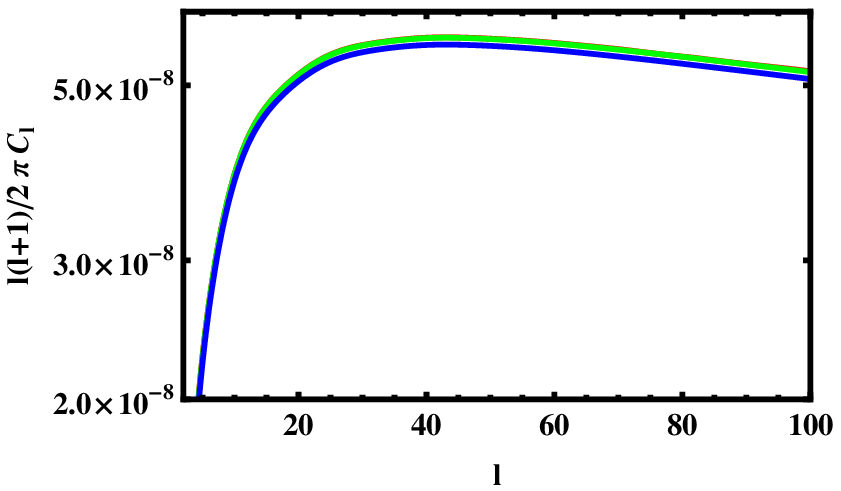,height=5cm}}
\caption{ \label{f:effwind} The effect of a window function on the density contribution (top panel), redshift space distortion
(middle panel) and lensing contribution (bottom panel).
We have chosen $z_S=0.1$ and $\De z_S=0$ (no window, top curve, red), $\De z_S=0.002$ (middle curve, green) and $\De z_S=0.01$ (bottom curve, blue).}
\vspace{-0.5cm}
\end{figure}

The lensing term scales like $\ell^4$ and is  in principle of the same order as the
density and redshift space distortion terms. However it is given by an integral of the form
(see Appendix~\ref{app:Cls})
$$\frac{\ell^2(\ell+1)^2}{r_S^2}\int\frac{dk}{k}T^2(k)
\left[\int_{0}^{r_S}d\la \frac{r_S-r}{r}j_\ell(kr)\right]^2$$
which does converge when integrated over $k$. It is therefore dominated at $k=\ell/r$. 
(We have used Limber's approximation~\cite{loverde} to evaluate this integral 
which we have tested numerically and found to be of excellent accuracy.)  The contribution of the 
lensing term becomes more important at larger source redshift for small $\ell$. But
it always remains subdominant in the transversal power spectrum. 
In the bottom panel of Fig.~\ref{f:tot} we plot the ratio between the new contributions, i.e lensing term plus potential term, and the total
angular power spectrum. We see that neglecting the new contributions for $z_S\le 1$ represents an error of no more than
0.1 percent, whereas for $z_S=3$ the error amounts to a few percent. Note that we do not include the Doppler terms in the new contributions since they appear already in the original Kaiser formula \cite{kaiser} (even though there the term from expansion $\propto \dot\HH/\HH^2$, 
which is of the same order for redshifts $z\ge 1$ is not considered). 
In Fig.~\ref{f:zl20} (top panel) we depict the redshift dependence of all the terms 
for a fixed value of $\ell=20$. The lensing and potential terms are both negative at small
redshift and become positive at large redshift. This is due to the fact that at small 
redshift the dominant contribution comes from their correlation with the density that is 
negative, whereas at large redshift the dominant contribution is their auto-correlation, 
$C_\ell^{LL}$, respectively $C_\ell^{\Psi\Psi}$. The bottom panel of Fig.~\ref{f:zl20} shows the ratio between the new contributions and the total angular power spectrum. The error induced by neglecting the new terms increases with redshift and it reaches a few percent at high redshift.

\subsection{The radial power spectrum}

\begin{figure}[H]
\centerline{\epsfig{figure=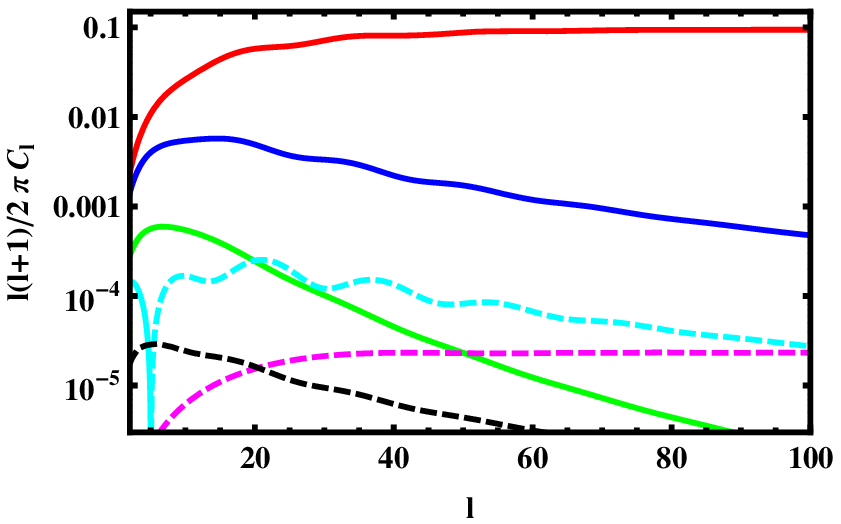,height=5cm}}
\centerline{\epsfig{figure=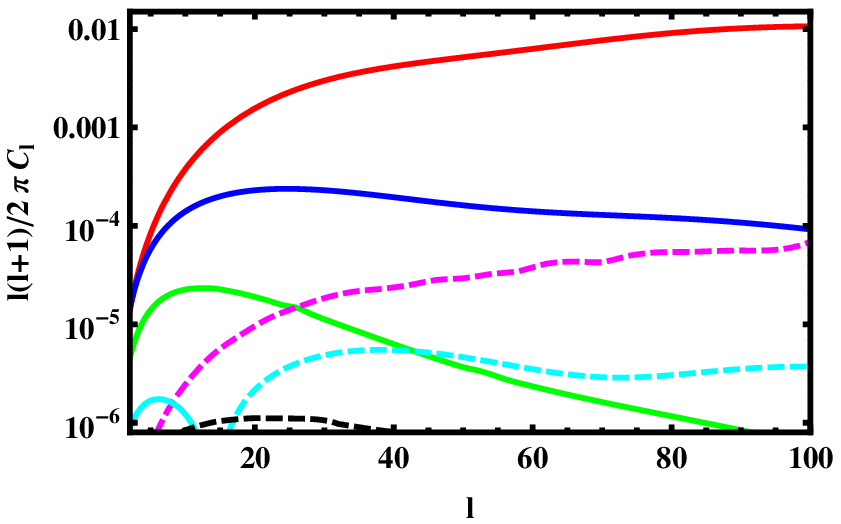,height=5cm}}
\centerline{\epsfig{figure=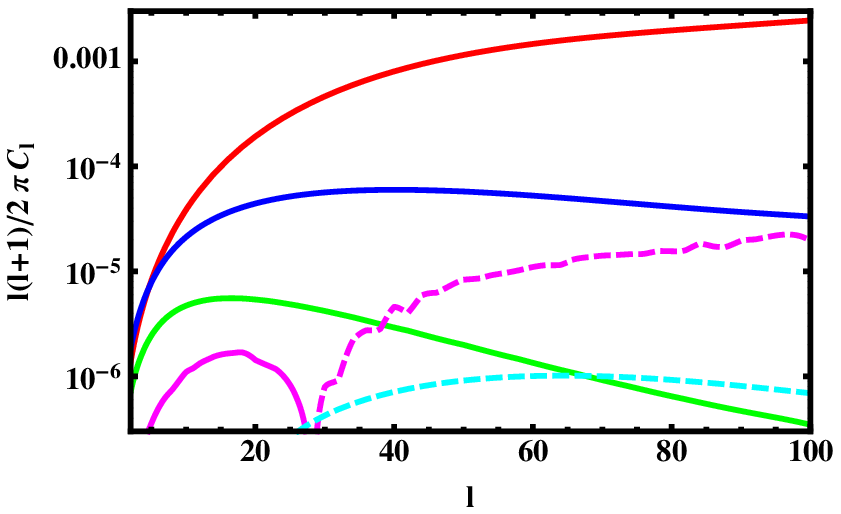,height=5cm}}
\centerline{\epsfig{figure=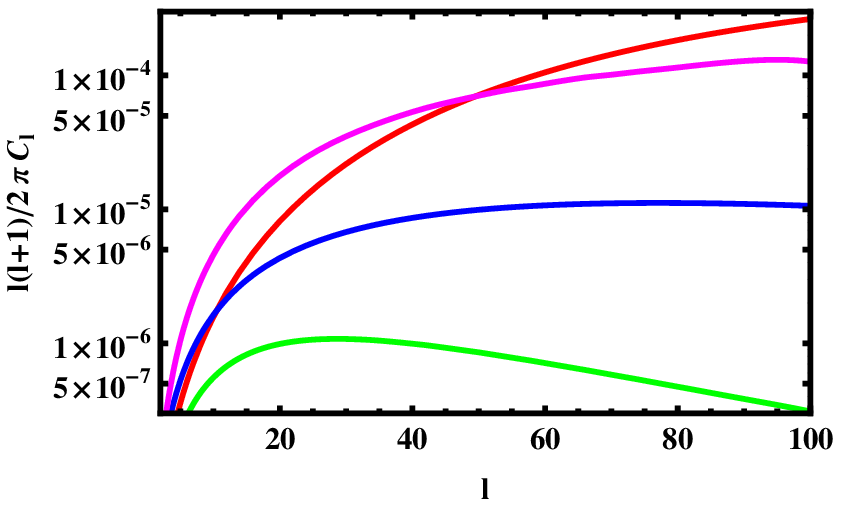,height=5cm}}
\caption{ \label{f:win} The effect of a window function with width $\De z_S=0.1z_S$ on the power spectrum $C_\ell(z_S)$
for redshifts (from top to bottom) 
$z_S=0.1,~0.5,~1$ and $3$. The different curves are: density (red), 
redshift space distortion (green), the correlation of density with redshift space distortion (blue), lensing (magenta), Doppler (cyan)
and gravitational potential (black), see Table~\ref{t:color}. Solid lines denote positive contributions whereas dashed lines denote negative contributions.}
\end{figure}

The results above give us the transversal power spectrum at fixed redshift. But of 
course there is also a radial 
power spectrum which correlates fluctuation at different distances from us.  This encodes different information and it is important to study them 
both. From the fact that the transfer function is not direction dependent, we infer that 
\be
\langle a_{\ell m}(z_S)a_{\ell' m'}(z_{S'})\rangle  = \de_{\ell,\ell'}\de_{m,m'}C_\ell(z_S,z_{S'}) \,.
\ee
Hence the radial power spectrum is given by
\be
C_\ell(z_S,z_{S'}) =  \frac{2A}{\pi}\int\frac{dk}{k}(kt_O)^{n-1} F_\ell(k,z_S)F^*_\ell(k,z_{S'}) \,.
\ee
 Here a interesting new phenomenon occurs: due to the fact that we evaluate 
$F_\ell(k,z_S)$ at different redshifts, we also evaluate the Bessel functions $j_\ell(kr_S)$ 
at different distances $r_S$. This leads to a suppression of the result due to oscillations, 
if the region in $k$-space where the integrand dominates has $kr _S> \ell$. 
As we discussed above, this is the case for the $k^2$--term of the density 
fluctuations and for the redshift space distortion, the terms which dominate the 
transversal power spectrum. These terms are therefore substantially suppressed in 
the radial power spectrum. All other terms have convergent integrals of 
$j^2_\ell(kr_S)$, already when neglecting the turnover of the 
transfer function, hence they are suppressed by powers of $\ell$ with respect to the lensing term. 
Therefore the lensing term dominates the radial power spectrum at low $\ell$.
This is precisely what one sees in Fig.~\ref{f:z01to3l20}, where the lensing term (magenta)
dominates for $z_{S'}$ significantly larger than $z_S$. As in Fig.~\ref{f:zl20},
at small redshift $z_S=0.1$ and $z_S=0.5$ the correlation density-lensing dominates 
(and is negative), whereas at large redshift $z_S=3$ the lensing-lensing term dominates. 
It is interesting to note how constant the lensing term remains while the density term and 
the redshift space distortion decay very rapidly with growing redshift difference. At $z_S=1$ 
the lensing-lensing
term and its correlation with the density are of the same order of magnitude which
explains the change of sign as $z_{S'}$ increases. Finally, at $z_S=0.1$ the Doppler term dominates over the standard term for some very specific values of $z_{S'}$. The first of them 
is actually the zero in the real space correlation function which e.g. at redshift $z_S=0.1$
corresponds to $\delta z = 0.011$.

\begin{figure}[!h]
\centerline{\epsfig{figure=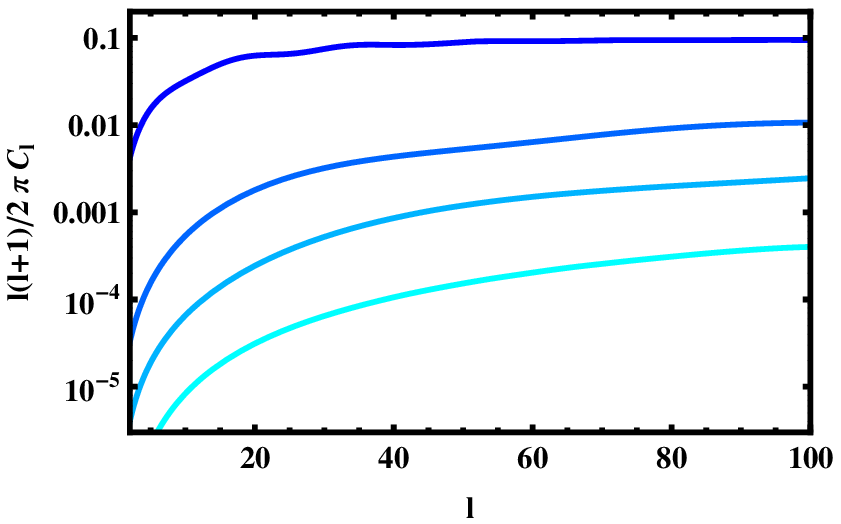,height=5cm}}
\centerline{\epsfig{figure=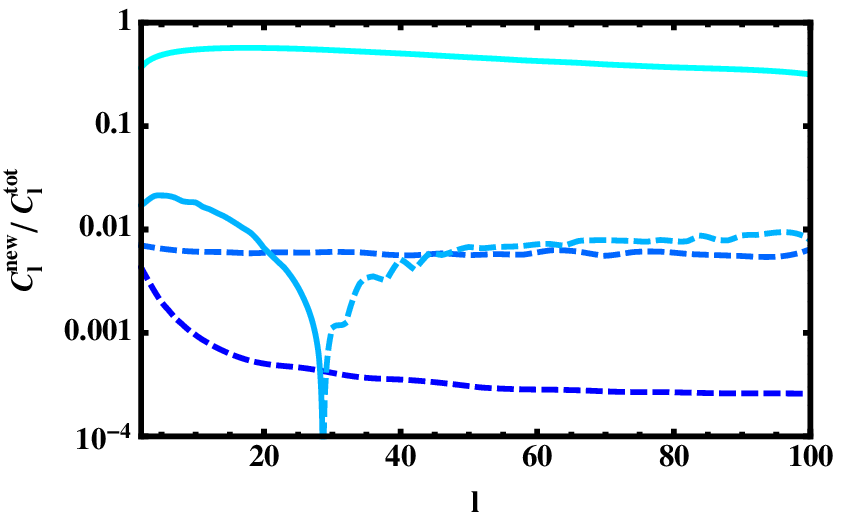,height=5cm}}
\caption{ \label{f:totwin} Top panel: The total power spectrum at redshifts (from top to bottom) $z_S=0.1$, $z_S=0.5$, 
$z_S=1$ and $z_S=3$ smeared by a window function with width $\De z_S =0.1z_S$.\\
Bottom panel: The ratio between the new contributions (lensing+potential) and the total angular power spectrum 
at (from top to bottom) $z_S=3$, $z_S=1$, $z_S=0.5$ and $z_S=0.1$. 
Solid lines denote positive contributions whereas dashed lines denote negative contributions.}
\end{figure}

\begin{figure}[!h]
\centerline{\epsfig{figure=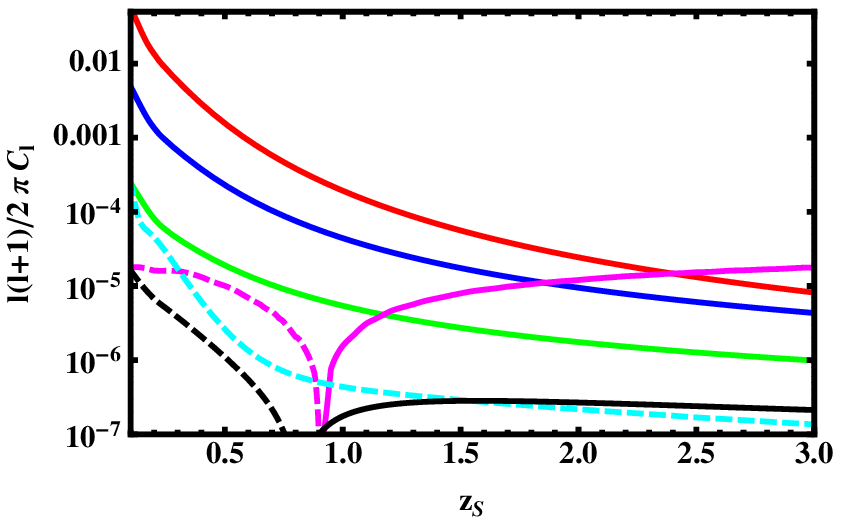,height=5cm}}
\centerline{\epsfig{figure=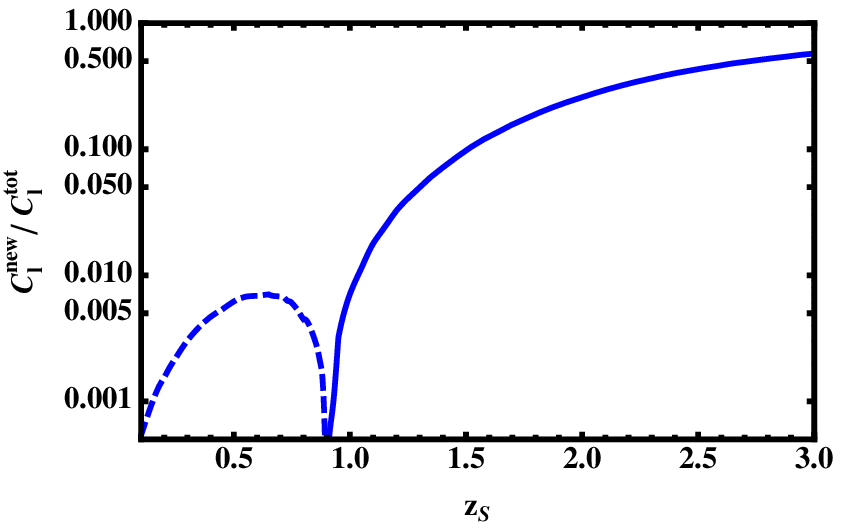,height=5cm}}
\caption{ \label{f:zwinl20} The various terms as a function of $z_S$ for fixed value of
$\ell=20$ and smeared by a window function with width $\De z_S =0.1z_S$: density (red), redshift space distortion (green), the correlation of density with redshift space distortion (blue), lensing (magenta), Doppler (cyan) and potential (black), 
see Table~\ref{t:color}.
Here the correlations between the lensing and Doppler and the lensing and potential are not negligible
at large $z_S$, and they are included in the Doppler (cyan), respectively potential (black) 
curves. Solid lines denote positive contributions, dashed lines denote negative contributions.\\
Bottom panel: The ratio between the new contributions (lensing+potential) and the total angular power spectrum,
smeared by a window function with width $\De z_S =0.1z_S$, plotted
as a function of $z_S$ for fixed value of $\ell=20$.}
\end{figure}

An alternative way to measure radial correlations is to introduce a window function $W(z,z')$ 
which corresponds to a smearing of fluctuations on scales smaller than some width $\De z_S$.
We use a Gaussian window around some mean redshift $z_S$ with width $\De z_S$. 
This suppresses power which comes from values of $k$ with $k\De r_S>\ell$ where
$\De r_S = r(z_S+\De z_S)-r(z_S)$. This is also a more realistic case since we can measure the galaxy 
distribution only in redshift bins of some finite width. Already a small width does 
substantially affect the resulting spectrum of the density, see Fig.~\ref{f:effwind} top panel, 
and the redshift space distortion (middle panel). As expected the lensing term is 
insensitive to this smearing (bottom panel).

In Fig.~\ref{f:win} we show the effect of a 10\% window on the different terms at different redshifts.
As before, the terms which we indicate by 'lensing term' 'Doppler term' and 'gravitational potential
contributions' in the figure are not only the corresponding term themselves but also 
their correlations with all other terms. If the latter dominate such a contribution can become
negative. For example the lensing contribution for $z_S=1$ changes sign at $\ell=28$. For
$\ell >28$ it is dominated by negative correlations with the density while for $\ell<28$ the
positive autocorrelation dominates. Since the power
from scales smaller than $k\De r_S$ is removed,  the power at $\ell$ truly
corresponds to that at $k = \ell/r(z)$ in the power spectrum. The 'wiggles'  in the 
density and in the velocity terms for $z_S=0.1$ are the baryon acoustic oscillations (BAOs),
the first of which appears at $\ell \simeq 15$ for $z_S=0.1$. They are also visible in the 
anti-correlation of the lensing term with the density for $z_S=0.5$ and $z_S=1$, but these 
terms are probably too small to be detected in real data.

In Fig.~\ref{f:totwin} (top panel) we show the total $C_\ell(z_S)$'s smeared with a 10\% window function.
Comparing it with Fig.~\ref{f:tot}, we mainly notice that the power is 
reduced significantly, by nearly 1.5 orders of magnitude. Furthermore, at $z_S=0.1$, the 
BAO's are clearly visible.  In the presence of a window, different terms can dominate
at different redshift and for different values of $\ell$. In the bottom panel of Fig. ~\ref{f:totwin}, we depict the ratio between
the new contributions and the total angular power spectrum. Neglecting the new contributions
induces an error of a few percent already at redshift 1, and this error increases to roughly 
50 percent at redshift 3. Note that this ratio depends strongly on the width of the window function,
and that a wider window would lead to a larger error. 

In Fig.~\ref{f:zwinl20} (top panel) we 
plot the different terms as a function of redshift, for fixed value of $\ell=20$. 
Contrary to Fig.~\ref{f:zl20}, where the lensing term always remains subdominant 
with respect to the density and redshift space distortion term, we see in Fig.~\ref{f:zwinl20} 
that for $z_S>2.4$ the lensing term dominates over the standard contribution. The redshift 
at which this dominance takes place depends of course on the chosen window function:
for larger $\De z_S$, the lensing term starts to dominate at smaller redshift. In the bottom panel
of Fig.~\ref{f:zwinl20} we show the ratio between the new contributions and the total angular power spectrum
as a function of redshift for $\ell=20$. From this figure we understand why in Fig.~\ref{f:totwin} (bottom panel), the ratio
at $z_S=1$ is not significantly larger than at $z_S=0.5$. The lensing contribution changes sign around $z_S=0.9$
and consequently it is still small at $z_S=1$. At a redshift of $z_S=1.5$, however the error 
induced by neglecting the new terms is already of the order of 10 percent.

\begin{figure}[!h]
\centerline{\epsfig{figure=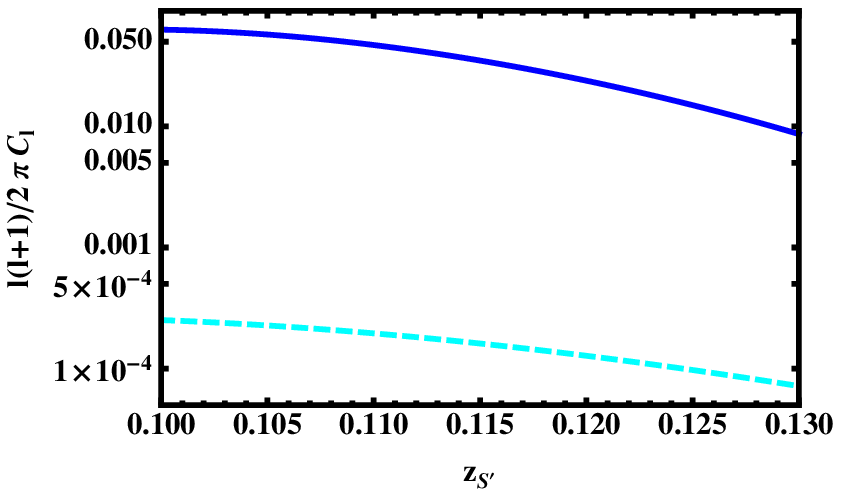,height=4.9cm}}
\centerline{\epsfig{figure=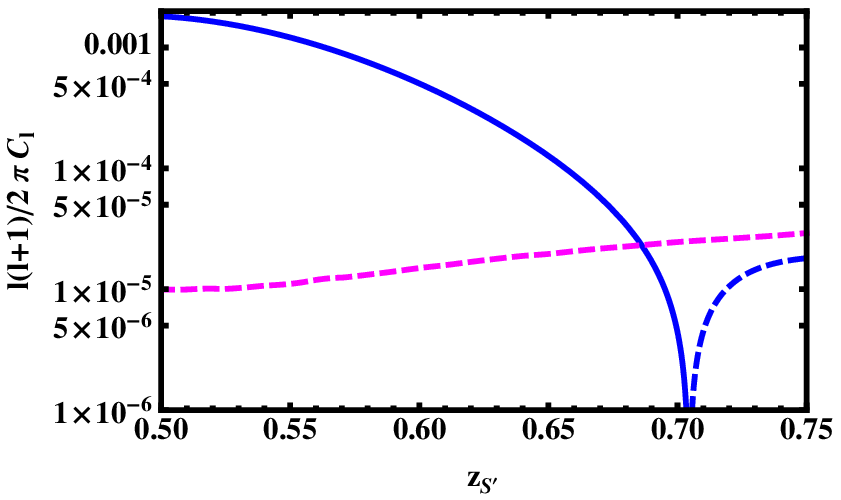,height=4.9cm}}
\centerline{\epsfig{figure=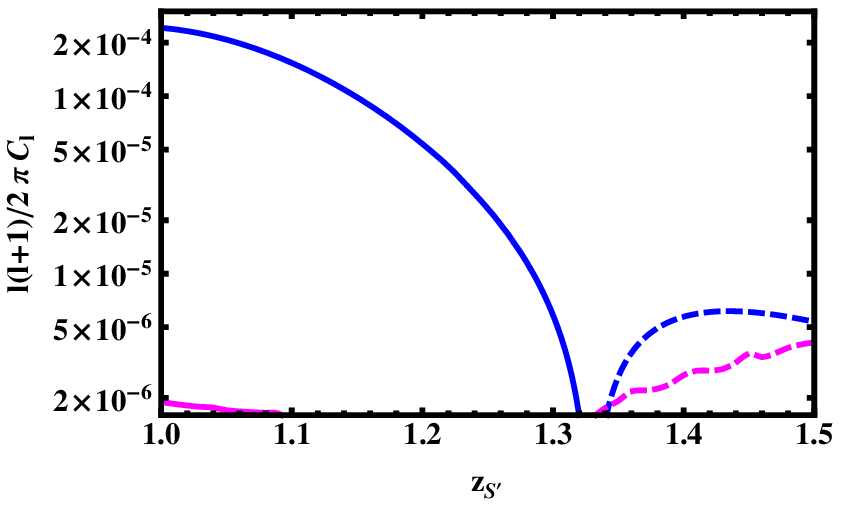,height=4.9cm}}
\centerline{\epsfig{figure=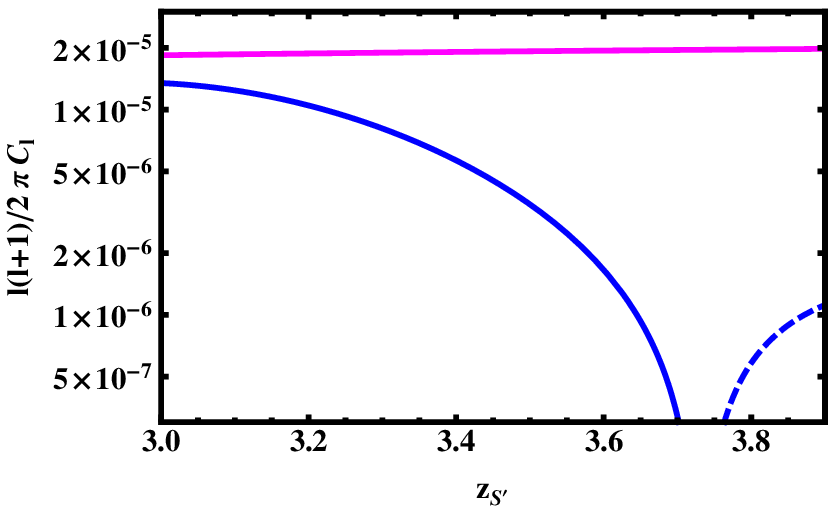,height=4.9cm}}
\caption{ \label{f:winz01to3l20} Cross-correlations between different redshift bins $C_\ell(z_S,z_{S'})$ at $\ell=20$ with a $10\%$ window function and plotted as a function of $z_S'$. From top to bottom $z_S=0.1,~0.5,~1$ and $3$.  Standard term, \ie $C^{DD}_\ell+C^{zz}_\ell+2C^{Dz}_\ell$ (blue), lensing (magenta), Doppler (cyan), potential (black), see Table~\ref{t:color}.
Solid lines denote positive contribution whereas dashed lines denote negative contributions.}
\end{figure}

In Fig.~\ref{f:winz01to3l20}, we show correlations between different redshifts bins (with a 
$10 \%$ window 
function),  for a fixed value of $\ell=20$. As in Fig.~\ref{f:z01to3l20} we see that the lensing 
term becomes dominant when the redshift separation between the bins increases. At large
redshift, $z_S=3$, the lensing
term is always dominant. The individual behaviour of each contribution is however quite 
different from Fig.~\ref{f:z01to3l20} which is due to the smearing introduced by the window 
function. Note that comparing the second panel in
Fig.~\ref{f:winz01to3l20} with the results in~\cite{Thomas}, we see that the redshift separation between their 4 different bins (their Fig.~13) is too small for the lensing contribution to be relevant. However, a similar measurement with one of the bins situated
around $z_S=0.7$ would already allow to detect the lensing contribution.

\begin{figure}[!ht]
\centerline{\epsfig{figure=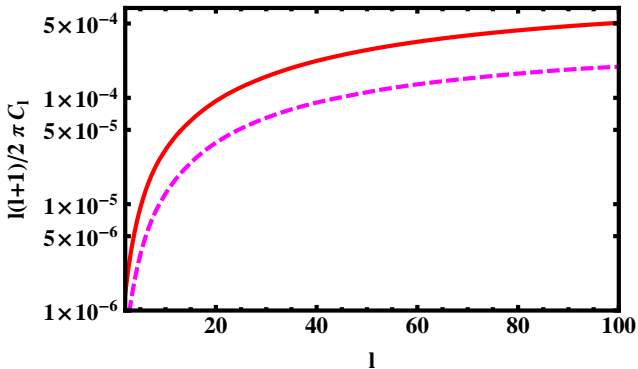,height=5cm}}
\caption{ \label{f:cintz} Integrated power spectrum with a flat distribution between $z_{\min}=0.1$ 
and $z_{\max}=2$ with Gaussian tails at both ends. The density term is plotted in red 
and the lensing term in magenta. Note that the lensing term is completely dominated 
by its anti-correlation with the density and hence is negative. }
\end{figure}

Finally, we plot in Fig.~\ref{f:cintz} the angular power spectrum integrated from the observer 
until a maximum redshift $z_{\max}$. This corresponds to the situation where the redshifts of 
individual galaxies is unknown but obeys a given redshift distribution. Consequently only 
the integrated spectrum can be measured. We assume a flat distribution of galaxies between 
$z_{\min}=0.1$ and $z_{\max}=2$ with Gaussian tails at both ends. In Fig.~\ref{f:cintz} we see 
that the only relevant contributions are the density and the lensing, more precisely the 
cross-correlation between the lensing and the density which is negative. The redshift 
space distortion contribution is, as expected, completely negligible when  the galaxy 
redshifts are unknown. The lensing contribution, however is  very relevant; it reduces the 
result by roughly $40 \%$ of the contribution from the density alone.

\section{Conclusions}

In this paper we have derived expressions for the transversal and radial galaxy
power spectra, $C_\ell(z_S)$ and $C_\ell(z_S,z_{S'})$, taking into account not only redshift 
space distortions, which have also been studied e.g. in~\cite{Thomas} but also all 
other relativistic effects to first order in perturbation theory. Within our accuracy
we are in reasonable agreement with the simulated results of Ref.~\cite{Thomas} 
(their Fig.~4) which analyzes the SDSS data taking into account redshift space distortion 
but not the other terms, {\em e.g.} the lensing, appearing in our formula (\ref{Dez}). 
They also take into account non-linearities in the matter power spectrum by using 
halofit~\cite{CAMB}. This enhances their results with respect to ours.

We have seen that by measuring $C_\ell(z_S,z_{S'})$ for different redshift differences and
different $\ell$'s we can measure different combinations of terms which depend on 
cosmological parameters in a variety of ways.  Otherwise, one may
measure the $C_\ell$'s smeared over a given redshift bin, $\De z_S$,
$$ C_\ell(z_S,\De z_S) = \int dz dz' W(z,z')C_\ell(z,z')$$
where $W$ is a window function centred at $z_S$ with width $\De z_S$. Without smearing,
the density contribution and the redshift space distortion always dominate. When 
smearing is included these terms are reduced and the lensing term can dominate.

The method outlined in this paper represents a very flexible new path to
estimate cosmological parameters and to test the consistency of the concordance model 
of cosmology. Of course, to do this we must master possible degeneracies not only
with biasing but also evolutionary effects  which have not been discussed in this work
and which may become relevant at redshift larger than 1, see~\cite{AA} for a discussion.
A detailed parameter estimation forecast e.g. for Euclid is left as a future project.

\FloatBarrier

\acknowledgments{We thank Anthony Lewis and Anthony Challinor who were accidentally
working on a very similar project~\cite{AA}. We compared our results with them
and although the derivation is different the analytical results completely agree.
Anthony Lewis also shared their numerical results with us. This
comparison helped us considerably, and we also do agree on the numerical part.
We  acknowledge useful discussions with Francis Bernardeau, Chiara Caprini, 
Chris Clarkson, Martin Kunz, Roy Maartens and Francesco Montanari. We thank the referee for useful
suggestions. RD is supported by the Swiss National Science Foundation. CB is supported by a Herchel Smith Postdoctoral Fellowship and
by King's College Cambridge.}

\appendix
\section{Some details of the derivations for $\De(\bn,z)$}\label{app:a}
We consider a perturbed Friedmann metric,
\bea
ds^2 &=&a^2(t) \Big(-(1+2A)dt^2 -2B_idtdx^i +  \\
&& +[(1+2H_L)\de_{ij}+ 2H_{Tij} + 2H_{ij}]dx^idx^j\Big)\nonumber
\eea
Here $H_{ij}$ is the transverse traceless gravitational wave term and $A$, $B_i$, 
$H_L$ and $H_{Tij}$ are scalar degrees of freedom, two of which can be
removed by gauge transformations. In Fourier space $B_i = -\hat k_i B$ and $H_{Tij} = 
(\hat k_i \hat k_j -\de_{ij}/3)H_T$. Often one uses longitudinal (or Newtonian) gauge with 
$B=H_T=0$  but we shall not use longitudinal gauge here. This is
useful if we want to determine whether a given expression is gauge invariant. 
In a generic gauge, the gauge invariant Bardeen potentials $\Phi$ and $\Psi$ are given 
by~\cite{Bardeen}
\bea
\Psi&\equiv&A+\frac{\HH}{k}B+\frac{1}{k}\dot{B}-\frac{\HH}{k^2}\dot{H}_T-\frac{1}{k^2}\ddot{H}_T\\
\Phi&\equiv&-H_L-\frac{1}{3}H_T+\frac{\HH}{k^2}\dot{H}_T-\frac{\HH}{k}B \, .
\eea
In longitudinal gauge they reduce to $A=\Psi$, $H_L=-\Phi$.

From this we easily obtain the following expressions for the redshift 
perturbation~\cite{myrev,mybook},
\bea\label{dez}
\frac{\de z}{1+z} &=&  
  -\Big[\big(H_L +\frac{1}{3}H_T + \bn\cdot\bV +
  \Phi+\Psi\big)(\bn,z)      \nonumber \\   &&   \qquad
 +  \int_{0}^{r_S}(\dot\Phi+\dot\Psi +\dot H_{ij}n^in^j)d\la \Big]
\eea
which leads to the density fluctuation in redshift space given in Eq.~(\ref{dez2}).
To determine the volume perturbation we have to compute the derivative,
\be
\label{eA:drdt_gi}
\frac{d\de r}{d\la}=-(\Phi+\Psi)+\frac{1}{k}\frac{dB}{d\la}+
\frac{1}{k^2}\left(\frac{d^2H_T}{d\la^2}-2\frac{d\dot{H}_T}{d\la} \right) 
-H_{ij}n^in^j \, .
\ee
Here we made use of the fact that 
the Fourier transforms of  $B_i$ and $H_{T ij}$ are respectively 
\bea
B_i(\bk,t)&=&-\frac{1}{k}\dd_iB\\
H_{T ij}(\bk,t)&=&\frac{1}{k^2}\dd_i\dd_jH_T+\frac{1}{3}\de_{ij}H_T .
\eea
Using  $\frac{dX}{d\la} = \dot X + n^i\dd_iX =  \dot X - \dd_rX$, we then obtain
\bea
B_in^i&=&-\frac{1}{k}\frac{dB}{d\la}+\frac{1}{k}\dot{B}  \nonumber\\
B_ie_{\theta}^i&=&-\frac{1}{kr}\dd_\theta B\nonumber\\
H_{T ij}n^in^j&=&\frac{1}{k^2}\left(\frac{d^2 H_T}{d\la^2}-2\frac{d\dot{H}_T}{d\la}+\ddot{H}_T \right) 
   +\frac{1}{3}H_T\nonumber\\
   &=& \frac{1}{k^2}\dd^2_rH_T +  \frac{1}{3}H_T\nonumber\\
H_{T ij}e_\theta^in^j&=&\frac{\dd_\theta}{k^2r}\left(\frac{d H_T}{d\la}-\dot{H}_T \right)+
  \frac{\dd_\theta }{(kr)^2}H_T .\nonumber
\eea
The angular volume perturbation is,
\bea
\left(\frac{\de v}{v}\right)_\Om \!\! &\equiv&  (\cot\theta+\dd_\theta) \de\theta +\dd_\varphi\de\varphi
 \nonumber \\  &=& 
\frac{-1}{r_S}\int_{0}^{r_S} d\la\frac{(r_S-r)}{r}\Delta_\Om(\Phi+\Psi) 
 - \frac{\Delta_\Om H_T(t_S)}{(kr_S)^2}  \nonumber \\
 &&  \qquad-\int_{0}^{r_S} d\la\left[\frac{(r_S-r)}{r_Sr}\Delta_\Om \left(H_{ij}n^in^j\right) +
 \right.  \nonumber \\
 && \qquad \left.  \frac{2}{r}\left(
 (\cot\theta +\dd_\theta)(H_{ij}n^ie_\theta^j) +   \right. \right.  \nonumber \\
 && \qquad \left. \left.  \frac{1}{\sin\theta}
   \dd_\varphi(H_{ij}n^ie_\varphi^j)\right)\right].
\label{eA:angle_gi}
\eea
Here $\Delta_\Omega$ denotes the angular part of the Laplacian. The second integral 
in Eq.~(\ref{eA:angle_gi}) is the contribution from gravitational waves
which we shall not discuss further in this work.

Putting it all together in Eq.~(\ref{e:volpert}) , using also Eq.~(\ref{dzdt_gi}) for
the derivative of the perturbed redshift, we obtain the volume perturbation at
fixed conformal time $t$ or fixed background redshift $\bar z$. However, we 
need to evaluate the volume fluctuation at a fixed, {\em observed} redshift which
is related to the latter by
\bea
 \left.\frac{\de v}{\bar v}\right|_z &=& \frac{\dd_z\bar v \de z + \de v(\bar z)}{\bar v} \\
  &=&  \frac{ \de v(\bar z)}{\bar v}  + \left(\frac{2}{r_S\HH} - 4 + \frac{\dot\HH}{\HH^2}\right)
  \frac{\de z}{1+z}  \nonumber \\   &=&
  3H_L -\bv\cdot\bn + \left(\frac{\de v}{v}\right)_\Om +\frac{2\de r}{r_S} - \frac{d\de r}{d\la} +
    \nonumber \\
 && 
  \frac{1}{\HH (1+z)} \frac{d\de z}{d\la} + \left(\frac{2}{r_S\HH} - 4 + \frac{\dot\HH}{\HH^2}\right)
  \frac{\de z}{1+z} \nonumber .
  \eea
  To simplify the expressions we combine the terms $(d^2H_T/d\la^2 -
  2d\dot H_T/d\la)/k^2$ of $\frac{d\de r}{d\la}$ and $2/(k^2r_S)dH_T/d\la$ of $2\de r/r$ with
  $\Delta_\Om H_T/(kr_S)^2$ of the angular volume perturbation to
 \bea
&&-\frac{1}{k^2}\left(\frac{d^2 H_T}{d\la^2} -2\frac{d\dot H_T}{d\la}-\frac{2}{r_S}\frac{dH_T}{d\la}
+\frac{\Delta_\Omega H_T}{r_S^2} \right)\nonumber\\
&& \qquad =-\frac{1}{k^2}\left(\Delta H_T -\ddot{H}_T -\frac{2}{r_S}\dot{H}_T\right)\nonumber\\
&& \qquad =H_T+\frac{1}{k^2}\left(\ddot{H}_T+ \frac{2}{r_S}\dot{H}_T\right) . \nonumber
\eea
 We also use the gauge invariant velocity potential~\cite{myrev,mybook}
\be
V\equiv{\rm v}-\frac{1}{k}\dot{H}_T
\ee
 so that
 \be
\bv\cdot\bn=-\frac{1}{k}n^i\dd_i v=\bV\cdot\bn-\frac{1}{k^2}n^i\dd_i\dot{H}_T
\ee
and the derivative of $\Phi$ along the light ray,

\bea
\frac{1}{\HH}\frac{d\Phi}{d\la}&=&-\frac{1}{\HH}\frac{d}{d\la}\left(H_L+\frac{H_T}{3}\right)
+\frac{1}{k^2}\left(\frac{d\dot{H}_T}{d\la}+\frac{\dot{\HH}}{\HH}\dot{H}_T \right)\nonumber\\
&&-\frac{1}{k}\left(\frac{dB}{d\la}+\frac{\dot{\HH}}{\HH}B \right)
\eea

With the help of these identities, the volume density perturbation reduces to the gauge invariant 
expression~(\ref{dev}), where the gravitational wave contribution is omitted for simplicity.

\onecolumngrid

\section{The contributions to the angular power spectrum \label{app:Cls}}
In this appendix we express certain contributions to the total $C_\ell's$ which are
of particular interest for the discussion given in the text. We use the transfer functions
for the concordance model given in eqs.~(\ref{TPhi}) to (\ref{TPsi}).

{\it Density fluctuation:}\\
Let us first consider the density term. The term of $T_D$ in (\ref{TD})
proportional to $k^2T_\Psi$  largely dominates the integral. Its contribution is
\be
C^{DD}_\ell(z_S)=\frac{2A}{\pi}\left(\frac{9}{10} \right)^2\frac{4a_S^2}{9\Om_m^2}\left(\frac{D_1(a_S)}{a_S} 
\right)^2 \int \frac{dk}{k}\left(\frac{k}{\HH_0} \right)^4T^2(k)j_\ell^2(kr_S) \,.
\ee
This integral only converges since $T(k)$ decays like $1/k^2$ for $k>k_{\rm eq}$, 
where $k_{\rm eq}$ is the (comoving) horizon scale at equal matter and radiation, 
see e.g.~\cite{mybook}. This integral is always dominated by the fluctuations on this scale, 
even at low $\ell\ll\ell_{\rm eq}(z) \simeq \pi k_{\rm eq}r(z_S) $.

{\it Redshift-space distortion:}\\
The term $T^2_V(k)\left(j_\ell''(kr_S)\right)^2$ coming from $\dd_r(\bV\cdot\bn)$ is the 
redshift space distortion. Since it is multiplied by $k/\HH$ its dominant contribution 
behaves like the density term and is of the same order
\be
C^{zz}_\ell(z_S)=\frac{2A}{\pi}\left(\frac{9}{10} \right)^2\frac{4a_S^2}{9\Om_m^2}
\left[\frac{D_1(a_S)}{a_S} +a_S\frac{d}{da_S}\left(\frac{D_1(a_S)}{a_S}\right)\right]^2
\int \frac{dk}{k}\left(\frac{k}{\HH_0} \right)^4T^2(k)j''^2_\ell(kr_S)\,.
\ee

{\it Cross-term density-redshift space distortion:}\\
Also this term is of the same order as the previous two and even dominates at low $\ell$.
\be
C^{Dz}_\ell(z_S)=-\frac{2A}{\pi}\left(\frac{9}{10} \right)^2\frac{4a_S^2}{9\Om_m^2}
\left[\frac{D_1(a_S)}{a_S} +a_S\frac{d}{da_S}
\left(\frac{D_1(a_S)}{a_S}\right)\right] \frac{D_1(a_S)}{a_S}\int \frac{dk}{k}\left(\frac{k}{\HH_0} \right)^4
T^2(k)j_\ell(kr_S)j''_\ell(kr_S)\,.
\ee

{\it Lensing:}\\
The lensing contribution is in principle also of the same order. But since it probes the 
power spectrum truly at $k\simeq \ell /r(z_S)$, it is largely subdominant at low $\ell$ if 
compared to the previous contributions.
\bea
C^{LL}_\ell(z_S)&=&\frac{8A}{\pi}\left(\frac{9}{10} \right)^2\ell^2(\ell+1)^2\frac{1}{r_S^2}\int\frac{dk}{k}T^2(k)
\left[\int_{0}^{r_S}d\la \frac{r_S-r}{r}\frac{D_1(a)}{a}j_\ell(kr)\right]^2\\
&=&4A\frac{\ell^2(\ell+1)^2}{(\ell+1/2)^3}\left(\frac{9}{10} \right)^2
\int_0^{r_S}\frac{dr}{r}\frac{(r_S-r)^2}{r_S^2}\left(\frac{D_1(r)}{a(r)}\right)^2
T^2\left(\frac{\ell+1/2}{r}\right)\,.\nonumber
\eea
In the last equality we have used Limber's approximation~\cite{loverde}
\be
\int_0^{y_S}dy f(y) J_\nu(y)=f(\nu)\theta(y_S-\nu)+O\left(\frac{1}{\nu^2}\right)\,.
\ee

{\it Cross-term density-lensing: }
\bea
C^{LD}_\ell(z_S)&=&-\frac{8A}{3\pi}\left(\frac{9}{10} \right)^2\frac{\ell(\ell+1)}{\Om_m r_S}
\int\frac{dk}{k}T^2(k)\left(\frac{k}{\HH_0} \right)^2 \frac{1}{1+z_S}\frac{D_1(a_S)}{a_S}j_\ell(kr_S)
\int_{0}^{r_S}d\la \frac{r_S-r}{r}\frac{D_1(a)}{a}j_\ell(kr)\\
&=&-\frac{8A}{3\pi}\sqrt{\frac{\pi}{2}}\left(\frac{9}{10} \right)^2\frac{\ell(\ell+1)\sqrt{\nu}}{\Om_m(1+z_S)}
\frac{D_1(a_S)}{a_S}\int_0^{r_S}\frac{dr}{r^3}T^2\left(\frac{\ell+1/2}{r}\right)\frac{r_S-r}{r_S}
\frac{D_1(r)}{a(r)}j_\ell\left( \frac{\nu r_S}{r} \right)\,.
\nonumber
\eea
The other terms are 

{\it Velocity:}
\bea
C^{VV}_\ell(z_S)&=&\frac{2A}{\pi}\left(\frac{9}{10} \right)^2\frac{4a_S^2}{9\Om_m^2}\left[\frac{D_1(a_S)}{a_S} +a_S\frac{d}{da_S}
\left(\frac{D_1(a_S)}{a_S}\right)\right]^2
\left(\frac{\HH_S}{\HH_0}\right)^2\left(\frac{\dot{\HH}_S}{\HH_S^2}+\frac{2}{r_S\HH_S}\right)^2\nonumber\\
&\cdot&\int \frac{dk}{k}\left(\frac{k}{\HH_0} \right)^2T^2(k)j'^2_\ell(kr_S)\,.
\eea

{\it Cross-term redshift space distortion-velocity: }
\bea
C^{zV}_\ell(z_S)&=&\frac{2A}{\pi}\left(\frac{9}{10} \right)^2\frac{4a_S^2}{9\Om_m^2}\left[\frac{D_1(a_S)}{a_S} +a_S\frac{d}{da_S}
\left(\frac{D_1(a_S)}{a_S}\right)\right]^2
\frac{\HH_S}{\HH_0}\left(\frac{\dot{\HH}_S}{\HH_S^2}+\frac{2}{r_S\HH_S}\right)\nonumber\\
&\cdot&\int \frac{dk}{k}\left(\frac{k}{\HH_0} \right)^3T^2(k)j'_\ell(kr_S)j''_\ell(kr_S)\,.
\eea

{\it Cross-term density-velocity: }

\bea
C^{DV}_\ell(z_S)&=&-\frac{2A}{\pi}\left(\frac{9}{10} \right)^2\frac{4a_S^2}{9\Om_m^2}\frac{D_1(a_S)}{a_S}
\left[\frac{D_1(a_S)}{a_S} +a_S\frac{d}{da_S}\left(\frac{D_1(a_S)}{a_S}\right)\right]
\frac{\HH_S}{\HH_0}\left(\frac{\dot{\HH}_S}{\HH_S^2}+\frac{2}{r_S\HH_S}\right)\nonumber\\
&\cdot&\int \frac{dk}{k}\left(\frac{k}{\HH_0} \right)^3T^2(k)j_\ell(kr_S)j'_\ell(kr_S)\,.
\eea

In the regimes investigated in this work the gravitational potential terms are always 
subdominant and we do not write them down explicitly.

\twocolumngrid

\end{document}